\documentclass[12pt]{article}
\usepackage{graphicx}
\usepackage{color}

\newcommand{\beq}{\begin{equation}}
\newcommand{\eeq}{\end{equation}}
\newcommand{\bea}{\begin{eqnarray}}
\newcommand{\eea}{\end{eqnarray}}

\newcommand{\gsim}{\lower.7ex\hbox{$
\;\stackrel{\textstyle>}{\sim}\;$}}
\newcommand{\lsim}{\lower.7ex\hbox{$
\;\stackrel{\textstyle<}{\sim}\;$}}

\newcommand{\eod}{\end{document}}

\begin{document}
\thispagestyle{empty}
\vspace*{-22mm}

\begin{flushright}
UND-HEP-15-BIG\hspace*{.08em}03\\
Version 7 \\

\end{flushright}
\vspace*{1.3mm}

\begin{center}
{\Large {\bf CP Asymmetries in Many-Body Final States in Beauty \& Charm Transitions}}

\vspace*{10mm}

{ I.I.~Bigi$^a$}\\
\vspace{7mm}
$^a$  {\sl Department of Physics, University of Notre Dame du Lac, Notre Dame, IN 46556, USA}\\

{\sl email addresses: ibigi@nd.edu} \\

\vspace*{10mm}

{\bf Abstract}
\vspace*{-1.5mm}
\\
\end{center}
Our community has focused on two-body final states in $B$ \& $D$ decays. 
The SM produces at least the leading source of CP violation in $B$ transitions; none has been established yet in charm decays. 
It is crucial to measure three- and four-body final states (FS)  
with accuracy and to compare with predictions based on refined theoretical tools.  
Correlations between different final states (FS) based on CPT invariance are often not obvious, 
how to apply them and where.  
We have to probe {\em regional} asymmetries and   
use refined parametrization of the CKM matrix.
One uses (broken) U- \& V-spin symmetries for spectroscopy. 
The situations with weak decays of hadrons 
are much more complex. The impact of strong re-scattering is large, and it connects 
U- \& V-spin symmetries. Drawing diagrams often does not mean we understand   
the underlying dynamics. I give a few comments about probing the decays of beauty \& charm baryons. 
I discuss the `strategies' more than the `tactics'.

\vspace{3mm}

\hrule

\tableofcontents
\vspace{5mm}

\hrule\vspace{5mm}

\section{Symmetries \& tools}
\label{History}

We have entered a novel era about probing heavy flavor dynamics, namely to connect "accuracy" and 
"correlations" on higher levels.  
The best fitted analyses are often not the best referees.  
I give a long introduction for several reasons. Often I disagree with statements in published papers about the  impact 
of U-spin symmetry; I will show a list of such articles at appropriate places.

(i) The Standard Model (SM) produces at least the leading source of the measured CP violation (CPV) in 
neutral kaons and $B$ transitions.

(ii) No CP asymmetry has been established yet in the decays of charm hadrons or baryons in general 
(except `our existence').

(iii) The neutral Higgs-like state has been found in the SM predicted mass region without sign 
of New Dynamics (ND) in its decays; on the other hand that has not been ruled out. 

(iv) Neutrino oscillations have been found with $\Delta m(\nu_i) \neq 0$ 
and three non-zero angles. There is a decent chance to 
find CP asymmetries there in the future despite the huge asymmetries in nuclei vs. 
anti-nuclei. 

(v) We have completely failed understanding  
the huge asymmetry in known matter vs. anti-matter in `our' universe.

Both on the experimental and theoretical sides we have measured CP asymmetries in mostly 
(pseudo-)two-body final states (FS) in the transitions of $K_L$ and $B$ mesons. 
However, we have to go beyond them, namely to measure "regional" asymmetries in 
three- \& four-body FS with accuracy. My main points are:  
those are {\em not} back-up of the informations we already get from two-body FS: they give us novel information about underlying dynamics, 
whether about non-perturbative forces of QCD 
and/or about ND. We have to discuss the impact of {\em re-scattering} with some details and how to test our tools. 
Of course, it needs much more work, but also tells us about fundamental forces. It is a true challenge to deal 
quantitatively with non-perturbative forces. 

Indirect CPV in the neutral mesons transitions ($K^0 - \bar K^0$, $B^0 -\bar B^0$, 
$B^0_s - \bar B^0_s$ and $D^0 - \bar D^0$) can be measured well in two-body FS, and 
it happened for $K^0$ and $B^0$ mesons already with good accuracy. 

Direct CP asymmetries in the decays of beauty and charm (\& strange) hadrons 
(\& $\tau$ leptons) 
need interferences between amplitudes with differences both in weak and strong phases 
\footnote{I do not include QED corrections; 
for now they cannot produce real impact for $B$, $D$ \& $\tau$.}. The first one comes from weak quark dynamics, 
which is not trivial, but still the easier part of our `job'.  
The second one depends on the impact of QCD. 
One can use `constituent' quarks that are models for spectroscopy of hadrons. 
Yet {\em current} quarks are based  
on real quantum field theories to describe rates of hadrons. There are subtle, but important differences between the 
worlds of hadrons and quarks usually named "duality".

In the SM with three families of quarks CP asymmetries are described by six triangles with different shapes, but all with the same area.   
Four of those can be probed directly about weak phases: 
\begin{enumerate} 
\item 
The triangle of $V_{ud}V^*_{ub}+V_{cd}V^*_{cb} + V_{td}V^*_{tb}  =0$ has been named the `golden' one 
in the previous millennium due to large CP asymmetries in $B^0$ \& $B^+$ transitions. 

\item 
{\em Indirect} CPV in $B_s \to \psi \phi, \psi f_0(960)$ is clearly smaller than in $B^0 \to \psi K_S$
due to $V_{us}V^*_{ub}+V_{cs}V^*_{cb} + V_{ts}V^*_{tb}  =0$, but still not very small:  
it is still possible around a few percent in $B^0_s$ transitions.  
\item 
SM produces very small CP asymmetries in singly Cabibbo suppressed amplitudes (SCS) of 
$D_{(s)}$  due to $V^*_{ud} V_{cd}+ V^*_{us} V_{cs}+V^*_{ub} V_{cb}=0 $ and much less for 
doubly Cabibbo suppressed (DCS) ones. The latter depends on weak coupling of $V^*_{cd}V_{us}$ 
in the SM. 
We know that the 2 x 2 sub-matrix of $V_{ud}, V_{us}, V_{cd},V_{cs}$  is not unitary: 
det$ [V_{2 \times 2}] \neq 1$; 
however it hardly produces measurable phase as discussed below.  

\item 
Super-tiny rates in $K^+ \to \pi^+ \nu \bar \nu$ 
\& $K_L \to \pi^0 \nu \bar \nu$ due to $V_{ud}V^*_{us}+V_{cd}V^*_{cs}+V_{td}V^*_{ts} =0$, 
where there is very good theoretical control of SM dynamics \cite{BURAS}. 
There is an experimental challenge to establish these two rates. 
\end{enumerate}
\begin{figure}[h!]
\begin{center}
\includegraphics[width=10cm]{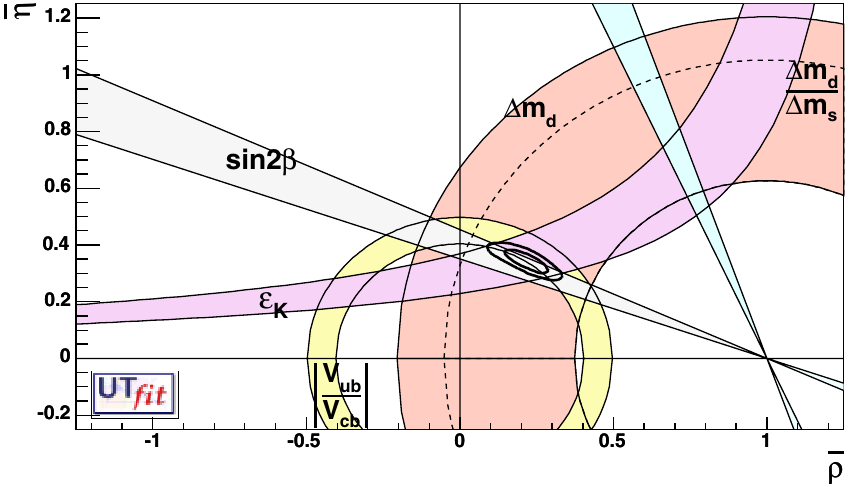} 
\end{center}
\caption{Correlations between other triangles ($\beta = \phi_1$)} 
\label{fig:TRIANGLES}
\end{figure}
One can look at the `golden' triangle with the limits from the measured values of $\epsilon_K$ 
and $\Delta M(B^0)/\Delta M(B^0_s)$ that connects, see Fig.\ref{fig:TRIANGLES}. Direct CPV has been established 
in $K_L \to \pi^+\pi^-$ vs. $K_L \to \pi^0\pi^0$. It is an interesting challenge for LQCD to understand 
in the future, whether ND could hide there so far \cite{NEWLQCD}. 
There are others, where we have not enough data to test them like 
for charm transitions and very rare decays of kaons. 
The correlations are crucial with other transitions. We are on the first step, but not even close to our goal, and 
we have to go for precision. 

In $\tau$ decays we expect simpler landscape in general. The SM predicts basically zero CP asymmetries in $\tau$ decays 
except in FS with the well measured $K^0 - \bar K^0$ oscillations (never mind where 
it comes from). One can use $\tau$ decays to calibrate the impact of low energy strong 
forces with accuracy and well tested theoretical tools like chiral symmetry. Yet surprises can happen.

The situation has changed significantly in the beginning of this millennium. We know that the 
SM produces at least the leading source of the measured non-zero CP asymmetries. 
We have to use refined theoretical (\& experimental) tools about hadronic forces. 
In this article I will focus on these items: 

(a) We have to probe regional CP asymmetries with accuracy in  three- \& four-body FS in the decays  
of charm and beauty hadrons. 
Beyond differences in averaged vs. regional asymmetries there are subtle questions: how do we 
{\em define} "regional" asymmetries and the impact of resonances in general, not only narrow ones? 

(b) It is important to measure {\em correlations} with FS from different decaying beauty \& charm (\& strange) hadrons. 
We have to work about these challenges with the best available tools for non-perturbative forces.

(c) It is not surprising to assume CPT invariance. In the worlds of quarks (\& gluons) one applies to classes of transitions, 
not only for total ones. For the worlds of hadrons it is much more subtle in general. Fundamental theories are formulated in the world of quarks. 

(d) Lipkin introduced U- \& V-spin symmetries as a good tool about 
{\em spectroscopy} in hadronic dynamics \cite{LIPKIN2}, in particular about light flavor 
baryons. The situation is much subtle 
about pions \& kaons; chiral symmetry solves the puzzles about their masses. 

(e) When one includes weak decays, the dynamics are more complex. For strange hadrons 
there is hardly a difference between exclusive and inclusive decays. However, there are 
large one for charm and huge one for beauty hadrons in non-leptonic ones. 

\noindent 
It seems to be popular to probe FS based on U-spin symmetry; below I give a list of papers about this item. 
I disagree with statements like "... model-independent 
\footnote{There are subtle differences between being `insensitive' vs. 
`independent' as discussed before; I prefer the word of `model-insensitive'.} 
relations are based on U-spin symmetry ...", in particular about its (semi-)quantitative impact. I will explain that below, 
why I disagree.

(f) When we discuss weak decays of hadrons, we have to think about the impact of strong 
{\em re-scattering} not only in principle, but also (semi-)quantitatively. 
It is very important to apply other tools 
in different levels with "judgment".

(g) Diagrams with quarks can show us the `road' to understand the underlying dynamics, 
but not always getting very close to our goal. 

(h) Of course, we follow the traditional steps: (i) models; (ii) model-insensitive analyses. However that is not the 
end of the road: we need another class of steps (iii), namely using refined tools to describe the data with good 
control. Very often the true theories do {\em not} give the best fitting of the data; their 
strengths are based on correlations with other transitions in a `network'. I focus on steps (ii) for now and talk about the strategies about  steps (iii) for the future. 

Before 1990's Lipkin discussed SCS $D^0 \to K^+K^-$ vs. 
$D^0 \to K^0 \bar K^0$  \cite{LIPK1}. Based on tree diagrams with four quarks one gets 
$\Gamma (D^0 \to K^0 \bar K^0) = 0$. To describe the existing 
data one needs strong "final state interaction" (FSI) that obviously comes from non-perturbative QCD forces. 
Obviously U-spin symmetry is not solid as isospin one by far. 
In particular, it was pointed out in Ref.\cite{HOLSTEIN} 
that it is not enough to say that a U-spin symmetry takes care of that problem -- there are dynamics effects in two-body FS, in particular by FSI; I agree. 
Lipkin mentioned that $\Gamma (D^0 \to K^0 K^-\pi^+)$ is {\em not} balanced by $\Gamma (D^0 \to K^0K^+\pi^-)$ based on 
U-spin symmetry breaking. 

Several of these items were mentioned in 1987 \cite{IBCHINA} and discussed in 1989 \cite{GOLDEN}, 
namely the impact of three-body FS. The review `The Physics of the B Factories' \cite{FINALBFACTORY} 
gives a broad context including charm \& $\tau$ decays. 

I `paint' the landscape of CP asymmetries about decays of beauty and charm hadrons first  
with averaged ones and later regional ones. It is easier to apply averaged strong phases to amplitudes for total rates. However the landscape is 
more complex for CP asymmetries, in particular for regional ones with three- and four-body FS -- like the large impact of resonances on probing CP asymmetries. 

(i) FSI and "re-scattering" basically refer to the same  strong dynamics mostly 
by non-perturbative QCD. Often authors use the words of FSI and "re-scattering" in somewhat different 
situations, but for my goals about strategies it makes hardly any difference. 
Therefore I use mostly the word of FSI, but sometimes I use "re-scattering" to make my point clearer, see 
Sects. \ref{EFFECT}, \ref{PENG}. 

(ii) There are three classes of diagrams:

\noindent 
Class I: Tree diagrams including perturbative QCD corrections are described with the usual symbol "$\to$". 

\noindent 
Class II: "Penguin" diagrams connect two quarks $Q$ \& $q$ have the same 
electric charge in their transitions, as I discuss below in Sect. \ref{PENG} in three {\em different} situations of dynamics. 
Even so, I use the same symbol "$\Longrightarrow$" just to remember the reader about these.

\noindent 
Class III: Weak annihilation (WA) \cite{WAWX} diagrams with the symbol "$\Rightarrow$". 
When one looks at with pseudo-scalar $Q\bar q$ with tree diagrams, 
one sees diagrams with weak annihilation or weak exchanges; however QCD corrections mix them, therefore 
one can use the word of WA in general.  
(Of course, the number of "colors" has impact.) On the other hand their impact is suppressed by 
chiral symmetry. The situation for baryons $Qq_1q_2$ is quite different: they are hardly suppressed by chiral symmetry -- 
but I use the same word "WA". 

I see no good reason why two-, three- \& four-body FS follow the same pattern; furthermore one should 
expect differences in beauty \& charm transitions. Finally I want to emphasize that we should not just `trust' diagrams. We have to check them due to "correlations" with other 
transitions and compare with other symmetries and tools.  
A well-known example is shown in the Fig. \ref{fig:TRIANGLES} in previous millenary. 
However the landscape is more complex now. I will discuss:   
diagrams vs. operators; penguin diagrams vs. FSI and  
measured rates \& asymmetries vs. (anti-)quark amplitudes with FSI. 
These items are obviously connected.

This paper is organized as follows: Sect. \ref{GENERAL} comments about 
the landscapes of inclusive vs. exclusive decays of charm \& beauty hadrons including 
CPT invariance and (broken) U- \& V-spin symmetry; Sects. \ref{BEAUTY} \& \ref{CHARM}
focus on beauty \& charm hadrons in very different landscapes; 
a summary is given in Sect. \ref{SUM}.

\section{Landscape of about beauty \& charm transitions} 
\label{GENERAL}

The connections of the worlds of hadrons and 
quarks (\& gluons) are often not straightforward. Actually there are also differences between the worlds of 
Hadrodynamics vs. HEP on the theoretical sides. Both of them discuss FSI, but in different landscapes and also where the term "duality" have some different meanings. 
Obviously I am biased there. 

For three-body FS we can use tested tools for probing Dalitz plots and 
better connect the worlds of hadrons and quarks. Those tools have been used before for strong 
transitions. Now we have to apply to weak decays of charm and beauty hadrons including low 
energy collisions of pseudo-scalar states -- in particular about the interferences for CP asymmetries. 
Those three-body FS are greatly affected both by {\em narrow} \& {\em broad} resonances; the broad ones, 
in particular about scalar ones, are {\em not} well described by Breit-Wigner (BW) parametrization.  
We have to probe four-body FS and go {\em beyond} their momenta. We can and should follow different `roads' to this `Rome'. 

I will give more detailed statements.  Not all of them are really novel, but often 
forgotten with their connections. We have to understand the information that the data give us 
as much as possible. 
There is also crucial difference between exclusive vs. inclusive rates \& asymmetries. Lipkin obviously knew about these differences and 
showed it \cite{LIPK1,LIPK2}. 

\subsection{Refined parametrization of the CKM matrix}
\label{CKM}

Dynamics of flavor violation in the SM world of hadrons are described as the first step by the CKM matrix. It is described by three angles and one weak phase 
(with three families). Most people use its 
parametrization going back to Wolfenstein that make its pattern obvious \cite{WOLFPARA}: 
with three parameters -- $A$, $\bar \rho$ and $\bar \eta$ -- assumed to be of the order of unitary and known $\lambda \simeq 0.225$ to be used for expansions 
in higher orders. It describes the data about flavor dynamics quite well including CP violation. 
There is only one subtle problem: data suggest that $|\bar \eta|$ and even more $|\bar \rho |$ are not of order unity:  
$|\bar \eta | \simeq 0.34$ and $|\bar \rho | \simeq 0.13$. It is surprising how 
this obvious pattern is so successful despite its disagreement with expected values 
of $\bar \eta$ and $\bar \rho$. 

Other parameterizations have been suggested for good reasons. One has been found specifically in Ref.\cite{AHN},  
when three parameters are truly of the order of unity 
($f \sim 0.75$, $\bar h \sim 1.35$ and $\delta_{\rm QM} \sim 90^o$), 
while the well-known $\lambda$ is not. The SM produces at least the leading source of CPV in measured $B$ transitions 
and predicts very small CPV in $D$ ones: 
\begin{eqnarray}
V_{\rm CKM}= \left(\footnotesize \begin{array}{ccc} 
V_{ud} & V_{us} & V_{ub} \\
V_{cd} & V_{cs} & V_{cb} \\
V_{td} & V_{ts} & V_{tb}
\end{array}\right)     = 
\; \; \; \; \;  \; \; \; \; \;  \; \; \; \; \; \; \; \; \; \;  \; \; \; \; \;  \; \; \; \; \; \; \; \; \; \; \; \; \; \; \; 
\; \; \; \; \; \; \; \; \; \; \;
&&   \\
=\left(\footnotesize
\begin{array}{ccc}
 1 - \frac{\lambda ^2}{2} - \frac{\lambda ^4}{8} - \frac{\lambda ^6}{16}, & \lambda , & 
 \bar h\lambda ^4 e^{-i\delta_{\rm QM}} , \\
 &&\\
 - \lambda + \frac{\lambda ^5}{2} f^2,  & 
 1 - \frac{\lambda ^2}{2}- \frac{\lambda ^4}{8}(1+ 4f^2) 
 -f \bar h \lambda^5e^{i\delta_{\rm QM}}  &
   f \lambda ^2 +  \bar h\lambda ^3 e^{-i\delta_{\rm QM}}   \\
    & +\frac{\lambda^6}{16}(4f^2 - 4 \bar h^2 -1  ) ,& -  \frac{\lambda ^5}{2} \bar h e^{-i\delta_{\rm QM}}, \\
    &&\\
 f \lambda ^3 ,&  
 -f \lambda ^2 -  \bar h\lambda ^3 e^{i\delta_{\rm QM}}  & 
 1 - \frac{\lambda ^4}{2} f^2 -f \bar h\lambda ^5 e^{-i\delta_{\rm QM}}  \\
 & +  \frac{\lambda ^4}{2} f + \frac{\lambda ^6}{8} f  ,
  &  -  \frac{\lambda ^6}{2}\bar h^2  \\
\end{array}
\right)
+ {\cal O}(\lambda ^7)
\label{MATRIX}
\end{eqnarray}
Thus the landscape of the CKM matrix is more subtle now: 
it is described by six triangles with six different patterns, but still with the same area:  
\bea
{\rm Triangle\; I.1:}&&V_{ud}V^*_{us} \; \; \;  [{\cal O}(\lambda )] + V_{cd}V^*_{cs} \;  \; \;  [{\cal O}(\lambda )] + 
 V_{td}V^*_{ts} \; \; \; [{\cal O}(\lambda ^{5\& 6} )] = 0   \\ 
{\rm Triangle\; I.2:}&& V^*_{ud}V_{cd} \; \; \;  [{\cal O}(\lambda )] + V^*_{us}V_{cs} \; \; \;  [{\cal O}(\lambda )] + 
V^*_{ub}V^*_{cb} \; \; \; [{\cal O}(\lambda ^{6 \& 7} )] = 0    \\
{\rm Triangle\; II.1:}&& V_{us}V^*_{ub} \; \; \;  [{\cal O}(\lambda ^5)] + V_{cs}V^*_{cb} \;  \; \;  [{\cal O}(\lambda ^{2 \& 3} )] + 
V_{ts}V^*_{tb} \; \; \; [{\cal O}(\lambda ^2  )] = 0   \\ 
{\rm Triangle\; II.2:}&& V^*_{cd}V_{td} \; \; \;  [{\cal O}(\lambda ^4 )] + V^*_{cs}V_{ts} \; \; \;  [{\cal O}(\lambda ^{2\& 3})] + 
V^*_{cb}V^*_{tb} \; \; \; [{\cal O}(\lambda ^{2 \& 3} )] = 0  \\
{\rm Triangle\; III.1:}&& V_{ud}V^*_{ub} \; \; \;  [{\cal O}(\lambda ^4)] + V_{cd}V^*_{cb} \;  \; \;  [{\cal O}(\lambda ^{3\& 4} )] + 
V_{td}V^*_{tb} \; \; \; [{\cal O}(\lambda ^3  )] = 0   \\ 
{\rm Triangle\; III.2:}&& V^*_{ud}V_{td} \; \; \;  [{\cal O}(\lambda ^3 )] + V^*_{us}V_{ts} \; \; \;  
[{\cal O}(\lambda ^{3\& 4})] + 
V^*_{ub}V^*_{tb} \; \; \; [{\cal O}(\lambda ^4 )] = 0 
\label{NEW2}
\eea  
I give two examples with different reasons. 
(a) In the Wolfenstein parametrization one gets one of large weak phases from $V_{td}$ leading to Im $V^*_{td} V_{ts} = - A^2 \lambda ^5 \eta$. Now one gets 
$V^*_{td} V_{ts} = - f \bar h \lambda ^6 {\rm sin}\delta_{QM}$ from $V_{ts}$. 
(b) The measured indirect CPV in $B^0$ decays -- $S(B^0 \to \psi K_S) = 0.676 \pm 0.021$   --  
is close the maximal value that the SM can produce, namely $S(B^0 \to \psi K_S) \sim 0.72$, which is not close to 100 \% \cite{BUZIOZ}. 

The real challenges come from strong 
dynamics. QCD is the only local theory that can describe that; however it depends how much 
we can trust our control over strong forces. To move forward we need more work -- and intelligent judgment.

\subsection{Theoretical tool kits for many-body FS}
\label{THTOOL}

Quark diagrams give us two-dimensional plots.  However FS with more than three hadrons 
cannot be described like that in general.  
The connections of quark diagrams with operators are subtle, in particular about local vs. non-local operators; the latter depends on long-distances FSI. 

(a) Measuring two-body FS gives one-dimensional observables from the rates and only numbers 
of CP asymmetries. Probing Dalitz plots for CP asymmetries give two-dimensional observables as we have seen already 
about $B$ decays.  
One applies amplitudes for FS with hadrons and resonances: 
$P \to h_1[h_2h_3] + h_2[h_1h_3] + h_3[h_1h_2] \to h_1h_2h_3$. I am not claiming 
that amplitudes of three-body are described by a sum of two-body FS perfectly. To be 
realistic it is enough for a long time.

(b) It is a good reason to state  
that the analyses are model-insensitive.  
We have to measure correlations with other data. Using averaged strong phases in three-body FS (or more) is the  
first step to understand dynamics. We have the tools to probe Dalitz plots about regional asymmetries. 
One uses model-insensitive and then uses tools that are checked 
due to correlations with other transitions as long as they are "acceptable"; 
the meaning of that depends. 

(c) One has to be realistic when probing four-body FS and uses the best tested tools to analyze 
one-dimensional asymmetries. We are at the beginning to that road to understand the 
underlying forces. 

(d) There are many strengths of lattice QCD, yet re-scattering is not one of those. 

There are connections with effective quark operators and hadronic transitions due to `duality' -- but they are subtle. As discussed in details in Ref. \cite{Nequ5}, one cannot 
compare only the FS using measured masses of hadrons and suggested ones for quarks -- it 
misses the crucial point of duality, namely the impact of {\em non}-perturbative forces.  

The situation of CP asymmetries in beauty \& charm hadrons give `wonderful challenges' for probing ND. At least we learn about the impact of FSI in the world of hadrons.

\subsubsection{Effective transition amplitudes including FSI}
\label{EFFECT}

Based on CPT invariance one can describe the amplitudes of hadrons following the history sketched above; it is given 
in Refs.\cite{1988BOOK,WOLFFSI,CICERONE} and 
in Sect. 4.10 of Ref.\cite{CPBOOK} with much more details:
\bea
T(P \to f) &=& e^{i\delta_f} \left[ T_f + 
\sum_{f \neq a_j}T_{a_j}iT^{\rm resc}_{a_jf}  \right] 
\label{CPTAMP1} 
\\
T(\bar P \to \bar f) &=& e^{i\delta_f} \left[ T^*_f +
\sum_{f \neq a_j}T^*_{a_j}iT^{\rm resc}_{a_jf}  \right]   \; ;  
\label{CPTAMP2}
\eea 
$T^{\rm resc}_{a_jf}$ describe FSI between $f$ and intermediate 
on-shell states $a_j$ that connect with this FS. It points out that $f$ are different from $a_j$, 
but are in the same classes of strong dynamics. In the world of quarks 
one describes $a_j = \bar q_j q_j$ and $f= \bar q_k q_k + {\rm pairs \; of \;} \bar q_l q_l$ 
with $q_{j,k,l} = u,d,s$. 
\footnote{In the worlds of Hadrodynamics one can use basically the same equations, 
where the actors are $\pi \pi \to \bar KK$ plus pairs of $\pi$ etc. One sees there are different cultures 
in Hadrodynamics vs. HEP.}

One gets regional CP asymmetries, not just averaged ones: 
\beq
\Delta \gamma (f) = |T(\bar P \to \bar f)|^2 - |T(P \to f)|^2 = 
4 \sum_{f \neq a_j} T^{\rm resc}_{a_jf} \, {\rm Im} T^*_f T_{a_j} \; ; 
\label{REGCPV}
\eeq
these FS $f$ consist of two-, three-, four-body etc. pseudo-scalars. 
In principle one can probe local asymmetries -- but one has to be realistic with finite data and  
a lack of `perfect' quantitative control of non-perturbative QCD. We need real theories about understanding 
of SM \& ND dynamics. We test our understanding of the information due to correlations with 
the data in "acceptable" ways. 
This statement is subtle, but crucial; I will discuss them in some details. 

CP asymmetries have to vanish upon summing over all such states $f$ 
using CPT invariance between subclasses of partial widths: 
\beq
\sum_{f} \Delta \gamma (f) =
4 \sum_{f}\sum_{f \neq a_j} T^{\rm resc}_{a_jf} {\rm Im} T_f^* T_{a_j} = 0 \; , 
\eeq
since $T^{\rm resc}_{a_jf}$ \& Im$T_f^* T_{a_j}$ are symmetric \& antisymmetric, 
respectively, in the indices $f$ \& $a_j$. 
 
It is one thing to draw quark diagrams by adding pair of 
$\bar q q$, but it is quite another thing to trust them. How can one connect the data 
about the decays with two-, three-,  four-body FS with the information about the 
underlying dynamics? We have to apply several theoretical tools there,  connect with others 
transitions and think about their limits. I will discuss U-spin symmetry \& its uncertainties in some details 
for good reasons and its connections with V-spin one. To make the important statement with somewhat 
different words: I see no reason why $T^{\rm resc}_{a_jf} $ do not connect U- \& V-spin transitions as pointed out just above. 
I will discuss below in some details for beauty \& charm decays.

There is a crucial point to understand that, namely the connection of the world of measured rates \& their 
asymmetries and the amplitudes one gets from the dynamics of quantum field theories as you can see 
from Eqs. (\ref{CPTAMP1},\ref{CPTAMP2}). 
It is not one-to-one as one sees by diagrams -- it is more subtle. 
FSI happen all the times and often with sizable impact. 
The question is: how much and where. We need help from other tools like chiral \& G-parity symmetries,  
dispersion relations etc.  
to understand the information given by the data; and  the item of "duality" \cite{BSU} 
comes in different situations. 

\subsubsection{`Painting' the landscape of diagrams including penguin ones}
\label{PENG}

Diagrams can be classified in three classes: 
(a) Some diagrams describe lifetimes of beauty hadrons etc. with {\em local} operators based on OPE and 
HQE \cite{BSU}. 
(b) Others discuss {\em short }distances forces like $\Delta \Gamma (B^0_s)$ (or hadronic jets) \cite{LENZMEM}. 
(c) Finally others try to deal with {\em long} distance dynamics like re-scattering about 
$\bar q_i q_i \leftrightarrow \bar q_j q_j$; actually it is for important for  
$\bar q_i q_i \to \bar q_j q_j + \bar q_kq_k \bar q_l q_l + ...$. 

Specifically I comment about penguin diagrams. 
Those are used to describe their impact on the decays of beauty \& charm hadrons, where FS consist mostly of many-body hadrons, 
not from two-body FS or even not (pseudo-)two-body ones. 
Therefore I use the word of `painting' the landscape of penguins:  
the situations are much more complex and also different for beauty and charm hadrons; furthermore I see no reason why two-, three- and four-body FS follow the 
same pattern. 

Penguin diagrams were introduced by M. Shifman, A. Vainshtein and V. Zakharow in 1975 
\cite{SVZ}  to explain the measured $\Delta I(3/2) \ll \Delta I(1/2)$ amplitudes in kaon decays and later predicted the direct  
CP violation $\epsilon^{\prime}/\epsilon_K \neq 0$. These are based on {\em local} operators with two-body FS, 
although they come from loop diagrams. 

{\em Inclusive} decays of {\em beauty} hadrons show impact of penguin diagrams in two classes of flavor changes -- 
$b \Longrightarrow s$ and $b\Longrightarrow d$ -- in a calculated way, but hardly for exclusive ones. 
One can add pairs of $\bar qq$ to penguin (or tree) diagrams and claims to produce the numbers of hadrons one wants. 
However to draw a diagram is one thing, 
but to describe the process is quite another thing 
\footnote{It was predicted 1500 years ago by Marinus, student of Proklos (known Neoplatonist philosopher influential on Western medieval philosophy as well as Islamic thought): 
"Only {\em being} good is one thing -- but good {\em doing} it is the other one!"}.  
The connections of penguin and tree diagrams with reality are often fuzzy as pointed out in 
Refs. \cite{1988BOOK,WOLFFSI,LENZMEM}.  
The name of `penguin' diagrams is often used in a very broad sense: 
$Q\bar q_a \to q (\bar q_i q_i + \bar q_i q_i \bar q_j q_j +...)\bar q_a$ following unitarity, 
where $Q$ and $q$ quarks carry the same charge with or without local operators. 
Of course, one should not hide theoretical uncertainties.   
Penguin diagrams give 
also imaginary part that one needs for FSI \cite{1988BOOK,CPBOOK}. 
Non-local penguin operator with internal charm lines can produce FSI: with $2m_c < m_b$ they can be 
{\em on-}shell and thus produce an imaginary part. However they give us pictorials, but not much more.In the world of hadrons one measures rates and asymmetries, 
while in the world of (anti-)quarks one produces amplitudes 
that have to include re-scattering, as pointed out above in Eq.(\ref{REGCPV}). That connection is far from trivial, 
and we are not yet at the end of the road to understand the real dynamics.

\subsubsection{Connect U- \& V-spin symmetries: spectroscopy vs. weak decays}
\label{USPINSYM}

The global $SU(3)_{fl}$ was introduced first
with its three $SU(2)_I$, $SU(2)_U$ and $SU(2)_V$, when quarks were seen mostly as a mathematical  tool to describe the spectroscopies of hadrons, not as real physical states 
\cite{LIPKIN2}. Therefore `constitute' quarks were used. 
It is easier to discuss the masses 
of baryons than those of the mesons, since the latter are greatly affected  by chiral symmetry. When one compares the masses of nucleons, $\Lambda$ and $\Xi$, 
one suggests the values of $u$ \& $d$ constituent 
quarks $\sim 330$ MeV and for $s$ one $\sim 500$ MeV . We have a better 
understanding of that due to mixing of $\langle 0| \bar u u|0\rangle$, $\langle 0|\bar d d|0\rangle$ 
between $\langle 0|\bar s s |0\rangle$ with scalar resonances that are not OZI suppressed \cite{DR}. 

It makes sense to use U-spin symmetry about {\em spectroscopy} of beauty \& charm hadrons. However the situation is much more complex in general: 
FSI have important impact on weak amplitudes in general and 
in particular  for CP asymmetries. Therefore we cannot ignore the correlations with V-spin symmetry.  
To say it differently: we cannot focus only on two-body FS and even more with only charged ones in weak transitions. 
As shown in Eqs.(\ref{CPTAMP1},\ref{CPTAMP2},
\ref{REGCPV}) intermediate two states strongly re-scatter into FS with two-, three-, four-, ... states either in the world of hadrons or quarks. 
The main point is very general: there are very different time scales of weak vs. strong forces. 
Therefore re-scattering makes the differences 
between U- \& V-spin symmetries very `fuzzy'; actually they are connected as pointed out above in Sect.\ref{EFFECT}.

The problem is to deal with FSI quantitatively. 
Obviously U-spin symmetry is sizably broken. One guess is 
$(M_K^2 - M^2_{\pi} ) < (M_K^2 + M^2_{\pi} )$.  
More refined ones are based on the item of `constitute' quarks and give 
$m_u^{\rm const} \sim m_d^{\rm const} < m_s^{\rm const}$. One can use that for models to predict 
{\em exclusive} decays, but with large theoretical uncertainties. 

In the world of quarks one describes mostly inclusive transitions. "Currents"  
quarks with $m_u < m_d << m_s$ are based on theory. I-, U- \& V-spin symmetries deal 
with $u \leftrightarrow d$, $d \leftrightarrow s$ \& $u \leftrightarrow s$. 
These three symmetries are obviously broken on different levels, and these violations are  
connected in the SM. The operators producing inclusive FS depend on their CKM 
parameters and the current quark masses involved there. However the real scale for inclusive 
decays is given by 
the impact of QCD, namely $\bar \Lambda \sim 0.7 - 1$ GeV as discussed many times
\footnote{For good reasons one uses different and smaller $\Lambda_{\rm QCD} \sim 0.1 - 0.3$ GeV for 
describing jets in collisions.}.  
Thus the violations of U- \& V-spin symmetries are small, and tiny for I-spin one. 
We can deal with inclusive rates and asymmetries of beauty and maybe charm hadrons using  
effective operators in the world of quarks. 

The connections with inclusive with exclusive hadronic rates are not obvious at least, in particular about quantitative ways. The violations of 
I-, U- \& V-spin symmetries in the measurable world of hadrons are expected to scale by the differences in pion and kaon masses, 
which are {\em not} small compared to $\bar \Lambda$ (or  $[m^2_K - m^2_{\pi}]/[m^2_K + m^2_{\pi}]$). 
This is even more crucial about direct CP violation and the impact of FSI on amplitudes. 
One reason is that suppressed decays in the world of hadrons consist with larger numbers of states 
in the FS, where FSI have great impact with opposite signs. 
Furthermore the worlds of 
hadrons are controlled by FSI due to {\em non}-perturbative QCD; they show the strongest impact on exclusive ones. 
It can be seen in the sum of exclusive ones in 
large ratios that go up and down much more sizably; I will show well-known examples of that below.  
However the rules are not yet well established how to connect the worlds of Hadrodynamics and HEP.

\subsection{Three-body FS in the decays of beauty \& charm hadrons} 
\label{DISPERSION}

We have to probe three-body FS with accuracy. The first step is to measure averaged CP asymmetries there and 
analyze about connections with two-body FS. 
Then one probes regional CP asymmetries using different technologies for Dalitz plots. 
Ratios of regional asymmetries do not depend on production asymmetries. One gets more observables to check experimental uncertainties. 
On the theory side one needs much more work, but check theoretical uncertainties about the impact of 
non-perturbative QCD and the impact of the existence of ND and its features. We have seen that FSI have large impact, 
in particular for suppressed decays of beauty and charm hadrons. 

There is a good reason to analyze data with model-insensitive ways as the second step. It does not mean there is only one road, 
actually there are three classes discussed. 
There is no `golden' tool, when one search for the impact of ND: one uses refined CKM matrix and 
probes Dalitz plots with different `roads' and compare their results. 
One puts them into small bin $i$, 
and customarily one measures `fractional asymmetries' 
$\Delta (i) \equiv \frac{N(i) - \bar N(i)}{N(i) + \bar N(i)}$ or  
`Miranda' procedure \cite{MIR1,MIR2,REIS} uses `significance' 
$S_{CP} (i) \equiv \frac{N(i) - \bar N(i)}{\sqrt{N(i) + \bar N(i)}}$ \footnote{Obviously I might be biased about the strengths of these tools.}; or 
one can also probe {\em un}binned Dalitz plots \cite{WILL}. 
These tools mentioned above do that in different ways and compare their results. 
At least it helps our thinking. 

It is not the final step. I want to emphasize the impact of systematical ones in several directions. 
Furthermore one has to think about correlations with other amplitudes; it needs much more 
work, but also gives us an `award'.  
BW parameterizations do not well describe the impact of 
scalar resonances 
like $f_0(500)/\sigma$ \& $K^*(800)/\kappa$ both in charm and beauty hadronic FS. 
Furthermore we have to discuss about the lists of resonances that are included with finite data. 
Then we cannot focus only on regional CP asymmetries `inside' narrow resonances 
with low masses, but also probe `outside' asymmetries. Obviously we can use chiral symmetry to probe 
the FS. There is a subtle tool, namely refined dispersion relations \cite{KUBIS,DR};  
they are based on data with low energy collisions of pions and kaons. In the end we learn a lot about underlying 
dynamics. It is true there are prices: (a) One needs much more data \& analyses. (b) We cannot use the same 
tool kit for several transitions, as you can see in the references \cite{KUBIS,DR}. The landscape is complex; however 
we should not give up. The basic elements for three-body FS: $\pi^+\pi^- \Longleftrightarrow \pi^0 \pi^0$, $K^+K^- \Longleftrightarrow \bar K^0 K^0$, 
$K^+K^- \Longleftrightarrow \pi^+ \pi^-$ \& $ K^-\pi^+ \Longleftrightarrow \bar K^0 \pi^0$.

For the two-dimensional  Dalitz plots we have the tools to probe them with a long history in 
strong dynamics. 
One needs larger amounts of data and experimental work, 
but they also deliver information about the existence of ND and its 
features.  In the end we have to agree after discussions; at least they tell us new lessons about strong forces.

\subsection{Four-body FS with different `roads' to ND}
\label{FOURGENERAL}

It was first pointed out that four-body FS can and should be probed in special situations, namely 
$B \to V_1V_2 \to h_1h_2h_3h_4$ with $V_i$ describing vector resonances  \cite{GERMAN}. It was 
realized that we have a general situations in the decays of heavy flavor hadrons with four-body FS \cite{RIOMANI}. 
I will discuss these four-body FS about $\Delta B \neq 0 \neq \Delta C$ below. 

Traditionally one compares {\em T-odd moments} of $H_Q \to h_1h_2h_3h_4$ vs. 
$\bar H_Q \to \bar h_1\bar h_2\bar h_3 \bar h_4$ in centre-of-mass frame:  
$C_T \equiv \vec p_1 \cdot (\vec p_2 \times \vec p_3)$, 
$\bar C_T \equiv \vec {\bar p_1} \cdot (\vec {\bar p_2} \times \vec {\bar p_3})$ leading to 
{\em T-odd} observables:
\beq
A_{T} \equiv \frac{\Gamma_{H_Q}(C_T >0) - \Gamma_{H_Q}(C_T <0)}
{\Gamma_{H_Q}(C_T >0) + \Gamma_{H_Q}(C_T <0)} \; , \; 
\bar A_{T} \equiv \frac{\Gamma_{\bar H_Q}(- \bar C_T >0) - 
\Gamma_{\bar H_Q}(- \bar C_T <0)}
{\Gamma_{\bar H_Q}(-\bar C_T >0) + \Gamma_{\bar H_Q}(- \bar C_T <0)}
\eeq
FSI can produce $A_T$, $\bar A_T$ $\neq 0$ without CPV. However 
\beq
a_{CPV}^{T-odd} \equiv \frac{1}{2} (A_T - \bar A_T ) 
\eeq
would establish CP asymmetry. With more data \& more thinking we might have some ideas about 
`better'  
values of $d>0$ that do not depend only on experimental reasons: 
\bea
\nonumber
A_{T} (d) &\equiv &\frac{\Gamma_{H_Q}(C_T > d) - \Gamma_{H_Q}(C_T <-d)}
{\Gamma_{H_Q}(C_T >d) + \Gamma_{H_Q}(C_T < -d)} \\ 
\bar A_{T} (d)&\equiv &\frac{\Gamma_{\bar H_Q}(- \bar C_T >d) - 
\Gamma_{\bar H_Q}(- \bar C_T <-d)}
{\Gamma_{\bar H_Q}(-\bar C_T >d) + \Gamma_{\bar H_Q}(- \bar C_T < -d)}
\eea 
However we cannot stop there -- we need one-dimensional observables (at least); furthermore 
we have to understand the reasons why different observables are used and compare their impact. 
I give two examples.
\begin{itemize}
\item
One can measure the angle $\phi$ between the 
planes of $h_1h_2$ and $h_3h_4$ and described its dependence 
\cite{CPBOOK, TAUD+}: 
\bea
\frac{d\Gamma}{d\phi} (H_Q \to h_1h_2h_3h_4) &=& \Gamma_1 {\rm cos^2}\phi + 
\Gamma_2 {\rm sin^2}\phi +\Gamma_3 {\rm cos}\phi {\rm sin}\phi
\\ 
\frac{d\Gamma}{d\phi} (\bar H_Q \to \bar h_1 \bar h_2 \bar h_3 \bar h_4) &=& \bar \Gamma_1 {\rm cos^2}\phi + \bar \Gamma_2 {\rm sin^2}\phi -\bar \Gamma_3 {\rm cos}\phi {\rm sin}\phi
\eea
The partial width for 
$H_Q[\bar H_Q] \to h_1h_2h_3h_4[\bar h_1\bar h_2\bar h_3 \bar h_4]$ is given by $\Gamma _{1,2} [\bar \Gamma _{1,2}]$: 
$\Gamma_1 \neq \bar \Gamma_1$ 
and/or  $\Gamma_2 \neq \bar \Gamma_2$ represents direct CPV in the partial width:
\beq
\Gamma (H_Q \to h_1h_2h_3h_4) = \frac{\pi}{2} (\Gamma_1 + \Gamma_2)  \; \; \; {\rm vs.} \; \; \; 
\Gamma (\bar H_Q \to \bar h_1\bar h_2 \bar h_3 \bar h_4) = \frac{\pi}{2} (\bar \Gamma_1 + \bar \Gamma_2)  
\eeq 
$\Gamma_3$ and $\bar \Gamma_3$ represent T {\em odd} correlations; however \cite{RIOMANI,CPBOOK}: 
$\Gamma_3 \neq \bar \Gamma_3  \; \; \to \; \; {\rm CPV}$.    
Integrated rates give $\Gamma_1+\Gamma_2$ vs. $\bar \Gamma_1 + \bar \Gamma_2$; the moments of 
integrated {\em forward-backward} asymmetry 
\beq
\langle A\rangle = 
\frac{\Gamma_3 - \bar \Gamma_3}{\pi(\Gamma_1+\Gamma_2+\bar \Gamma_1+\bar \Gamma_2)}
\eeq
gives information about CPV. When one has enough data to do that, one could disentangle $\Gamma_1$ vs. $\bar \Gamma_1$ and 
$\Gamma_2$ vs. $\bar \Gamma_2$ by tracking the distribution in $\phi$. If there is a {\em production} asymmetry, it gives global 
$\Gamma_1 = c \bar \Gamma_1$, $\Gamma _s = c \bar \Gamma _2$ 
and $\Gamma_3 = - c \bar \Gamma_3$ with {\em global} $c \neq 1$. 

\item 
We have learnt from  the history of 
$K_L \to \pi^+\pi^- \gamma ^*  \to \pi^+\pi^- e^+e^-$, 
where Seghal \cite{SEGHAL1,SEGHAL2} really predicted CPV there around 14 \% based on 
$\epsilon_K \simeq 0.002$. Of course, the landscapes of $\Delta S \neq 0$ and $\Delta B\neq 0 \neq \Delta C$ are quite different 
for several reasons; I mention only one now: $\Delta S \neq 0$ amplitudes are described with local operator, but not for the others. 

It helps to discuss that situation in more details with unit vectors: 
\beq
\vec n _{\pi} = \frac{\vec p_+ \times \vec p_-}{|\vec p_+ \times \vec p_-|} \; , \;
\vec n _{l} = \frac{\vec k_+ \times \vec k_-}{|\vec k_+ \times \vec k_-|}  \; , \;
\vec z = \frac{\vec p_+ + \vec p_-}{|\vec p_+ + \vec p_-|} \\
\eeq 
\bea
{\rm sin} \phi = ( \vec n _{\pi}  \times \vec n _{l} ) \cdot \vec z \; [CP=-,T=-] &,& 
{\rm cos} \phi = \vec n _{\pi}  \cdot \vec n _{l} \; [CP=+,T=+]\\
\frac{d\Gamma}{d\phi} &\sim & 1 -(Z _3\,  {\rm cos} 2\phi  + Z_1\, {\rm sin}2\phi)
\eea
Then one measures asymmetry in the moments:
\beq
{\cal A}_{\phi} = 
\frac{(\int_0^{\pi/2} - \int_{\pi/2}^{\pi}+\int_{\pi}^{3\pi/2}-\int_{3\pi/2}^{2\pi})\frac{d\Gamma}{\phi}}
{(\int_0^{\pi/2} + \int_{\pi/2}^{\pi}+\int_{\pi}^{3\pi/2}+\int_{3\pi/2}^{2\pi}) \frac{d\Gamma}{\phi}}
\eeq
There is an obvious reason to probe only the angle between the two 
$\pi^+\pi^-$ \& $e^+e^-$ planes. It is based on $K_L \to \pi^+\pi^- \gamma ^*$ 
or $K^0 \to \pi^+\pi^- \gamma ^*$ vs. $\bar K^0 \to \pi^+ \pi^- \gamma^*$. 

Here the situations are more complex for several reasons; therefore I use:
\bea
\nonumber 
\frac{d}{d\phi} \Gamma (H_Q \to h_1h_2h_3h_4) &=& |c_Q|^2 - 
[b_Q\,  {\rm cos} 2\phi  + a_Q\, {\rm sin}2\phi] = \\
&=&|c_Q|^2 - [b_Q\,  (2 {\rm cos}^2 \phi  -1) + 2a_Q\, {\rm sin}\phi \, {\rm cos}\phi]  \\
\nonumber
\frac{d}{d\phi} \Gamma (\bar H_Q \to \bar h_1\bar h_2\bar h_3\bar h_4) &=& 
|\bar c_Q|^2 - [\bar b_Q\,  {\rm cos} 2\phi  - \bar a_Q\, {\rm sin}2\phi ]= \\  
&=&|\bar c_Q|^2 - [\bar b_Q\,  (2 {\rm cos}^2 \phi  -1) - 2\bar a_Q\, {\rm sin}\phi \, {\rm cos}\phi]
\eea
Obviously the landscapes are complex 
\beq
\Gamma (H_Q \to h_1h_2h_3h_4) = |c_Q|^2 \; \; \;  {\rm vs.} \; \; \; 
\Gamma (\bar H_Q \to \bar h_1\bar h_2 \bar h_3\bar h_4) = |\bar c_Q|^2  
\eeq
For these moments one gets: 
\beq
\langle A_{\rm CPV}^Q \rangle = \frac{2(a_Q - \bar a_Q)}{|c_Q|^2+|\bar c_Q|^2}  \; ; 
\eeq
i.e., no impact from $b_Q$ \& $\bar b_Q$ terms. 

Furthermore one wants to probe semi-regional asymmetries like:
\beq
A_{\rm CPV}^Q|_e^f =
\frac{\int _e^f d\phi \frac{d\Gamma}{d\phi}- \int_e^f d\phi\frac{d\bar \Gamma}{d\phi}}
{\int _e^f d\phi\frac{d\Gamma}{d\phi}+ \int_e^f d\phi\frac{d\bar \Gamma}{d\phi}}
\eeq
where $b_Q$ and $\bar b_Q$ contribute. Again, the main point is not to choose which gives the 
best fitting one, but has deeper reasons. I will discuss more specifically later. 
\end{itemize}
These examples are correct from the general theoretical bases. However, some 
deal better with experimental uncertainties, cuts and/or probe the impact of ND; also the true underlying dynamics 
do not always -- actually often not -- produce the best fitting of the data. 
It is crucial to use CPT invariance as a tool for correlations with other transitions. 
It is important to probe 
semi-regional asymmetries and compare results with different tools, but also with more thinking. 
Somewhat similar statements can be seen in \cite{GDYG}.

\subsection{Short resume}
\label{SS}

My points are about CP asymmetries in the weak decays of charm and beauty hadrons:  
\begin{itemize}
\item
First one probes two-body FS for non-leptonic transitions. However one has to go beyond that, since many-body 
FS give us much more information about the underlying dynamics. I see no reason, why 
many-bodies FS follow the same pattern. Actually present data show different patterns, as I show below. 

\item
I suggest there are the same classes I, II \& III for amplitudes for 
two-, three- \& four-body FS. 
However the goals are quite different, namely leading sources 
for $\Delta C \neq 0$ or close to them, but non-leading ones for $\Delta B \neq 0$. 

\item
Indirect CPV is mostly best measured in time dependent rates with two-body FS.

\item 
CPT invariance produces much more equalities in the masses \& widths of particles and 
anti-particles: equalities of sub-classes of decays are defined by FSI due to strong dynamics, 
see Eqs. (\ref{CPTAMP1}, \ref{CPTAMP2}, \ref{REGCPV}). There are correlations between two-, three-, 
four-body FS that often are not obvious.

\item
A `popular' tool is used to connect different FS, namely (broken) U-spin symmetry. 
However we have to deal with more complex situations: 
\begin{itemize}
\item 
One can apply U-spin symmetry to discuss hadronic {\em spectroscopy} with decent uncertainties. 
However the landscapes have changed significantly, when one includes {\em weak} dynamics. 
Re-scattering connect U- \& V-spin symmetries. Below I will discuss statements given in a list of papers based on U-spin symmetry  
and explain why I disagree. Those papers had just ignored previous published papers without 
explaining why. 

\item
Even in the world of quark diagrams one cannot ignore re-scattering 
$\bar q_i q_i \to \bar q_j q_j + \bar q_j q_j \bar q_k q_k ...$.  
Furthermore we cannot describe that with local operators in general -- we need long distance forces. 

\item
There are crucial differences between "inclusive" vs. "exclusive" decays: we learn from the data about the 
impact of underlying both weak and strong dynamics. One 
can use expansions of $(m_s - m_d)/m_c$ for "inclusive" one about U-spin symmetry and likewise for 
$(m_s - m_u)/m_c$ about V-spin one. Both of them are small. However, the landscape is changing 
sizably, when we discuss about "exclusive" ones.

\item
We cannot focus only on (quasi-)two-body FS like $2\pi$, $\pi \rho$, $2\rho$, $\bar K K$, $\bar K K^*$ etc. 
\end{itemize} 
\item
Production asymmetries in $pp$ collisions obviously affect in particular charm and beauty {\em baryons}  decays. 
On the other hand, class III amplitudes can contribute sizably here, since WA \cite{WAWX} diagrams of beauty \& charm baryons are not suppressed mostly 
by chiral symmetry in opposite what happens for meson decays. 
It would be a great achievement by the LHCb collab. to establish CPV in heavy flavor baryons' decays 
no matter where it comes from. I will comments mostly about $\Lambda_c^+ \to p K^+\pi^-$.

\item 
One has to include scalar resonances that usually are broad ones and are not well described by 
BW parameterization. With finite data one has to think and discuss about the lists of the 
resonances included. 

\item
The best fitted analyses do often not give the best understanding of the dynamics.

\item
I see no reason, why the impact of FSI should follow the same pattern. 

\end{itemize}
We have to discuss correlations with other transitions.  
Good judgment understands where theories (even theorists) help.

To describe the decays of beauty \& charm hadrons it remembers me of the Austrian saying: 
"It is the same -- only different!" 
The same classes of tools can be used -- differently.

\section{$\Delta B \neq 0$  forces}
\label{BEAUTY}

The measured lifetimes of $B^0$ and $B_s^0$ are the same within 2\% as a 
sign of U-spin symmetry about {\em inclusive} transitions; likewise for the inclusive 
semi-leptonic decays. Furthermore they agree the theoretical predictions within those 
uncertainties.  

However the landscapes for {\em exclusive} non-leptonic decays are different, as you see in PDG2015: 
\bea
{\rm BR}(B^0 \to K^+\pi^-) &=& (1.96 \pm 0.05) \cdot 10^{-5}  \\
{\rm BR}(B^0_s \to K^-\pi^+) &=& (0.55 \pm 0.06) \cdot 10^{-5} 
\eea
In principle those branching ratios are `expected', since penguin diagrams $b \Longrightarrow d$ are more suppressed 
for $B^0_s$ transitions. On the other hand $b \Longrightarrow d$ produces large weak phases. Therefore we are 
not surprised by the data: 
\beq
A_{CP} (B^0 \to K^+\pi^-) = - 0.082 \pm 0.006 \; , \; 
A_{CP} (B^0_s \to K^+\pi^-) = + 0.263 \pm 0.035
\label{ACPTWOBODY}
\eeq
Based on U-spin symmetry it was suggested in Refs.\cite{GR2000,LIPK2,IMBLON11} to probe  
\beq
\Delta = \frac{A_{CP} (B^0 \to K^+\pi^-)}{A_{CP} (B^0_s \to K^+\pi^-)} + 
\frac{{\rm BR}(B^0_s \to K^-\pi^+)}{{\rm BR}(B^0 \to K^+\pi^-)} \frac{\tau _d}{\tau _s}=0
\eeq
Actually it depends only on two-body FS with charged kaon \& pion: 
\beq
\Delta = \frac{A_{CP} (B^0 \to K^+\pi^-)}{A_{CP} (B^0_s \to K^+\pi^-)} + 
\frac{\Gamma (B^0_s \to K^-\pi^+)}{\Gamma (B^0 \to K^+\pi^-)}
\eeq 
Precent data from LHCb give\cite{LHCbPRL110}:
\beq
\Delta = -0.02 \pm 0.05 \pm 0.04 \; . 
\label{PRL110}
\eeq
It is  not clear what the data tell us: is it $\Delta \simeq 0$ -- the strength of U-spin symmetry -- or 
$\Delta \sim -0.1$ -- the impact of re-scattering? To say it with  different words about important questions. 
On which scale is U-spin symmetry broken? 
Can one focus only on charged two-body FS? We cannot ignore re-scattering, namely 
$K^- \pi^+ \Leftrightarrow \bar K^0 \pi^0$ etc. etc.; more importantly, as I said just above: 
$(s \bar u)(u \bar d) \to \bar K \pi^{\prime}s$, $\bar K K \bar K \pi^{\prime}s$ and even 
$(s \bar q)(q\bar d) \to \bar K \pi^{\prime}s$, $\bar K K \bar K \pi^{\prime}s$ with $q=u,d,s$.  
One expects larger impact of penguin diagrams for $B^0$ vs. $B^0_s$ transitions, at least in the SM: 
$\Gamma (B^0 \to \bar K \pi^{\prime}s/\bar K K \bar K \pi^{\prime}s) > \Gamma (B^0_s \to \bar K \pi^{\prime}s/\bar K K \bar K \pi^{\prime}s)$. 
We know that U- \& V-spin symmetries are connected with strong forces, as discussed in general in Sects. \ref{EFFECT}, \ref{PENG} \& \ref{USPINSYM}. 

It has been stated already in the Abstract of the Ref.\cite{IMBLON11}: "pure-penguin decay $B^0_s \to \bar K^0 K^0 \bar K^0$ is intriguing ...";  
I have to disagree. The connections of quark diagrams with hadronic amplitudes is "complex". Below I will discuss in general about 
$B^{\pm}$ transitions. Here I talk about this special example: one looks at tree operators $b \to u (d\bar u)$ \& 
$b \to d (u \bar u)$ and penguin one $b\Longrightarrow d$ and embanked into the $B^0_s = [\bar b s]$ wave function. 
The authors of Ref.\cite{IMBLON11} said: when one describes three-body FS in SM suppressed $B^0_s$ 
decays, one needs only one pair of $\bar qq$ for tree diagrams and two pairs of $\bar q_1q_1\bar q_2 q_2$. 
Therefore tree diagrams produce only $B_s^0 \to \bar K^0 K^+K^-$, but not 
$B^0_s \to \bar K^0 K^0 \bar K^0$. I disagree with such statements, 
since re-scattering with strong forces produces $\bar u u \to \bar d d$ or 
$\bar u u \to \bar s s$ and also tree diagrams give $B^0_s \to \bar K^0 K^0 \bar K^0$ in principle, although we cannot give a semi-quantitative production. 
On the other hand, we have two examples for CP asymmetries in three-body FS of suppressed $B^{\pm}$ decays, where we found very interesting 
data, as I discuss below.

\subsection{CP asymmetries in $B^{\pm}$ with CPT invariance}

LHCb data show that CKM suppressed $B$ decays mostly populate the boundaries of Dalitz plots. 
At the qualitative level one should not been surprised. The `centers' are not really empty. 
CP asymmetries come from interferences; therefore one expects large regional ones -- 
but how much and where? The impact of broad resonances might be best seen in asymmetries 
rather than rates.

When one discusses CKM suppressed $B^-$ amplitudes in the SM, one starts with local operators  based on  left-handed currents 
$[\bar u \gamma_{\mu}(1- \gamma_5) b][\bar q \gamma_{\mu}(1- \gamma_5) u ]$   
(with $q = s,d$) leading to renormalized operators 
\beq
O_{\pm} \propto \{ [\bar u \gamma_{\mu}(1- \gamma_5) b][\bar q \gamma_{\mu}(1- \gamma_5) u ] \pm 
 [\bar q \gamma_{\mu}(1- \gamma_5) b][\bar u \gamma_{\mu}(1- \gamma_5) u ]  \} \; .
\eeq
A local penguin operator $O_{Peng}$ $b \Longrightarrow$ $q$ ($q=s,d$) connects a left-handed current with a vector one 
due to QCD. Thus one gets for the $B^-$ amplitude for 
$c_+ \langle u\bar u s \bar u  | O_{+}|B^- \rangle$ + $c_- \langle u\bar u s \bar u   | O_{-}| B^-\rangle $ 
+ $c_P \langle u\bar u s \bar u  | O_{Peng}|B^- \rangle$. 
CPT invariance gives: 
\bea
\Gamma (B^- \to \bar K +X_{S=0})  &=& \Gamma (B^+ \to  K +\bar X_{S=0}) \\
\Gamma (B^- \to X^{\prime}_{S=0})  &=& \Gamma (B^+ \to  \bar X^{\prime}_{S=0}) 
\eea
The $X$, $\bar X$, $X^{\prime}$ \& $\bar X^{\prime}$ in the FS include pairs of $\bar K K$. 
Duality tells us: 
\bea
\Gamma ([b\bar u] \to u\bar u s \bar u) &\simeq& \Gamma (B^- \to \bar K +X_{S=0}) \\
\Gamma ([b\bar u] \to u\bar u d \bar u) &\simeq& \Gamma (B^- \to  X^{\prime}_{S=0})
\eea
Unitary tell us that we can add pairs of $\bar q_i q_i$ to quark diagrams due to FSI. 
We can{\em not} trust diagrams 
even semi-quantitatively to connect $u \bar us \bar q_i q_i \bar u$ [$q_i=u, d, s$] 
with only $\bar K \pi \pi$ \& $\bar K \bar K K$ or $u \bar u d  \bar q_i q_i \bar u$ with 
$\pi \pi \pi$ \& $\pi \bar KK$ including neutral pions \& kaons to cancel asymmetry. 
What about hadronic FS $K^-\pi^+\pi^- \pi^0$, $K^-K^+K^-\pi^0$, $\bar K^0\pi^-\pi^+\pi^-$, 
$\bar K^0 K^-K^+\pi^-$? Or $4\pi$ or $\pi \pi \bar K K$? 
Duality connects the worlds of (anti-)quarks with the one 
of hadrons, namely with two- \& many-bodies FS; to be realistic we have to probe "regional" CP asymmetries 
with three- \& four-body FS.

\subsubsection{$B^{\pm} \to K^{\pm}\pi^+\pi^-$ vs. $B^{\pm} \to K^{\pm}K^+K^-$}

The data of CKM suppressed $B^+$ decays show no surprising rates in PDG2015:
\bea
\nonumber
{\rm BR}(B^+ \to K^+\pi^-\pi^+) &=&(5.10 \pm 0.29 ) \cdot 10^{-5}  
\label{Pen1}
\\  
{\rm BR}(B^+ \to K^+K^-K^+) &=&(3.37 \pm 0.22 ) \cdot 10^{-5} \; .
\label{Pen2}
\eea
Operators $O_+$ \& $O_-$ give the same large weak phase from $V_{ub}$, while the penguin operator gives 
zero weak phase. $O_{\rm Peng}$ contributes to the branching ratios. Penguin diagrams  
give pictorials for FSI. However, they do not allow us to calculate their impact on exclusive transitions beyond 
have-waving arguments. 

LHCb data show averaged CP asymmetries \cite{LHCb028}: 
\bea
\nonumber
\Delta A_{CP}(B^{\pm} \to K^{\pm} \pi^+\pi^-) &=&  
+0.032 \pm 0.008_{\rm stat} \pm 0.004_{\rm syst}
\pm 0.007_{\psi K^{\pm}}  
\label{SUPP1} 
\\
\Delta A_{CP}(B^{\pm} \to K^{\pm} K^+K^-) &=&   
- 0.043 \pm 0.009_{\rm stat} \pm 0.003_{\rm syst}
\pm 0.007_{\psi K^{\pm}} 
\label{SUPP2}
\eea 
with 2.8 $\sigma$ \& 3.7 $\sigma$ from zero. 
The sizes of these averaged asymmetries are not really surprising (unless one thinks more about 
interferences on these Dalitz plots); however it does not mean that we 
could really predict them. It is very 
interesting that they come with opposite sign due to CPT invariance.  

LHCb data show regional CP asymmetries \cite{LHCb028,ERNEST}: 
\bea 
\nonumber 
A_{CP}(B^{\pm} \to K^{\pm} \pi^+\pi^-)|_{\rm regional} &=& + 0.678 \pm 0.078_{\rm stat} 
\pm 0.032_{\rm syst}
\pm 0.007_{\psi K^{\pm}} 
\label{SUPP3} 
\\
A_{CP}(B^{\pm} \to K^{\pm} K^+K^-)|_{\rm regional} &=& - 0.226 \pm 0.020_{\rm stat} 
\pm 0.004_{\rm syst} \pm 0.007_{\psi K^{\pm}} \; .
\label{SUPP4}
\eea 
Regional CP asymmetries are defined by the LHCb collaboration: positive asymmetry at low 
$m_{\pi ^+\pi ^-}$ just below $m_{\rho^0}$; negative asymmetry both at low and high $m_{K^+K^-}$ values. 
It should be noted the opposite signs in Eqs.(\ref{SUPP1},\ref{SUPP3}). One needs more data about regional CP asymmetries -- 
but also more thinking and better theoretical tools for strong FSI. One expects larger regional 
CP asymmetries here --  but where and so large? Can it show the impact of broad resonances like $f_0(500)/\sigma$ and 
$K^*(800)/\kappa$? At least they give us highly non-trivial lessons 
about non-perturbative QCD.  We have to wait from Belle II to probe 
$B^{\pm} \to K^{\pm}\pi^0\pi^0$/$K^{\pm}\bar K^0 K^0$ 
(\& $B^{\pm} \to K^{\pm}\eta^{(\prime)}\pi^0$).

\subsubsection{$B^{\pm} \to \pi^{\pm}\pi^+\pi^-$ vs. $B^{\pm} \to \pi^{\pm}K^+K^-$}

Again the data of even more CKM suppressed $B^+$ decays show no surprising rates 
\bea
\nonumber 
{\rm BR}(B^+ \to \pi^+\pi^-\pi^+) &=& (1.52 \pm 0.14 ) \cdot 10^{-5} 
\label{Pen3} 
\\  
{\rm BR}(B^+ \to \pi^+K^-K^+) &=& (0.50 \pm 0.07 ) \cdot 10^{-5} \; .
\label{Pen4}
\eea
One expects smaller rates of these FS based on the `experience' from $\bar B_d \to K^-\pi^+$ 
vs. $\bar B_d \to \pi^+\pi^-$ ones; indeed it is true given for the data. 
While the amplitudes of $O_{\pm}$ operators are less suppressed above ones, 
$b \Longrightarrow d$ penguin diagram is more suppressed than $b\Longrightarrow s$ ones.
  
Data show {\em larger} CP asymmetries as discussed above in Eqs.(\ref{SUPP2},\ref{SUPP4}) \cite{IRINA} 
(again with the opposite signs):
\bea
\nonumber 
\Delta A_{CP}(B^{\pm} \to \pi^{\pm}  \pi^+\pi^-) &=&  
+0.117 \pm 0.021_{\rm stat} \pm 0.009_{\rm syst}
\pm 0.007_{\psi K^{\pm}}  
\label{SUPP5} 
\\
\Delta A_{CP}(B^{\pm} \to \pi^{\pm} K^+K^-) &=&   
- 0.141 \pm 0.040_{\rm stat} \pm 0.018_{\rm syst}
\pm 0.007_{\psi K^{\pm}}  \; . 
\label{SUPP6}
\eea 
However comparing CP asymmetries  show the surprising impact of penguin diagrams:   
in the SM one gets amplitudes $T(b \Longrightarrow d)$; it also produces weak phase with $V_{td}$ on the same {\em level} as $V_{ub}$ in the SM. 
Again: re-scattering happens and affects sizably transitions -- however where and how much?  

Again CP asymmetries focus on small regions in the Dalitz plots \cite{IRINA,ERNEST}.  
\bea 
\nonumber 
\Delta A_{CP}(B^{\pm} \to \pi^{\pm} \pi^+\pi^-)|_{\rm regional} &=&  
+0.584 \pm 0.082_{\rm stat} \pm 0.027_{\rm syst}
\pm 0.007_{\psi K^{\pm}}  
\label{SUPP7} \\
\Delta A_{CP}(B^{\pm} \to \pi^{\pm} K^+K^-)|_{\rm regional} &=&   
- 0.648 \pm 0.070_{\rm stat} \pm 0.013_{\rm syst}
\pm 0.007_{\psi K^{\pm}} \; .
\label{SUPP8}
\eea 
It should be noted also their signs in 
Eqs.(\ref{SUPP6},\ref{SUPP7}). Again, does it show the impact of 
broad scalar resonances like $f_0(500)/\sigma$ and/or $K^*(800)/\kappa$?

\subsubsection{Comparing with the literature for three-body FS}

There is a large literature about three-body FS in suppressed decays of $B^{\pm}$ and sizable ones for CP asymmetries 
including Dalitz plots. Obviously I agree we need to probe suppressed $B$ decays with three-body FS with accuracy in general; 
however, I disagree with several statements giving in three articles \cite{LOR,BGR,DONG}. 
Later I will discuss an important point, namely the connections with the $B$ transitions with $D$ \& $\tau$ ones. 
\begin{itemize}
\item
In those articles no reference was given about earlier papers, like Refs.\cite{MIR1,MIR2,BUZIOZ}; in two of them simulation was given 
for the possible impact of re-scattering. In \cite{MIR1} the connection of $B^{\pm} \to 3\pi$,$\pi\bar KK$, $K\pi\pi$ \& $K\bar K K$. 
The importance of CPT invariance was emphasized in \cite{MIR2}; in \cite{BUZIOZ} it was said that CPT invariance is `usable'. 
It was pointed out that U- \& V-spin symmetries are connected by strong forces. 

\item
One can compare $B^{\pm} \to K^{\pm} \pi^+\pi^-$, $K^{\pm}\pi^0\pi^0$, $K^{\pm} K^+K^-$, $K^{\pm} \bar K^0 K^0$.
In the future one wants to measure FS including $\eta^{(\prime)}$ and even think to include "constituent" gluons in their wave functions.

\item 
The impact of FSI was discussed in general \cite{1988BOOK,WOLFFSI,CICERONE,CPBOOK}. 

\item 
For probing weak decays of beauty mesons we have entered a new era, where we know that the SM gives 
basically the leading source of CP asymmetries. Therefore we need a refined parametrization of the 
CKM matrix for beauty (\& charm) hadrons as pointed out above. 

\item
I have said before that we should not forget the impact of CPT invariance in principle, but in charm 
transitions it is usable; we have learnt it is also usable for three- \& four-body FS of beauty hadrons, as 
I said in \cite{BUZIOZ}. For example, the data describe a surprising simple landscape, namely we find that 
$A_{CP}(B^+ \to K^+\pi^+\pi^-) \sim - A_{CP}(B^+ \to K^+K^+K^-)$ and $A_{CP}(B^+ \to \pi^+\pi^+\pi^-) \sim - A_{CP}(B^+ \to \pi^+K^+K^-)$ 
and the regional asymmetries are large. 

\item 
In my view it is not enough to fit the data in the best road; we have to use dispersion relations to 
understand the underlying dynamics as long as our analyses give us good results. 
The tools of dispersion relations have a successful \& long history \cite{DR,KUBIS} as 
mentioned in Sect.\ref{DISPERSION}. It shows the connection of Hadrodynamics \& HEP; so far it has been 
applied to favored transitions of charm mesons.

\end{itemize}

\subsection{Triple-product asymmetries}

BaBar  \cite{PHIKSTARBABAR}, Belle \cite{PHIKSTARBelle} and LHCb \cite{LHCb2014-005} have measured 
polarization amplitudes of $B^0 \to \phi K^*(892)$; the first two have measured also with other resonances. Their data 
have shown impact of re-scatterings, but have found no sign for CPV there. For example, LHCb analyses give 
$\Delta _{CP} = + 0.015 \pm 0.032 \pm 0.005$. There is a puzzle: 

\noindent 
(a) Large direct CPV has been established $\sim 10\; \& \; 20$ \% in $B^0 \to K^+\pi^-$ \& $B^0_s \to \pi^+ K^-$, see Eq.(\ref{ACPTWOBODY}). 

\noindent 
(b) Three-body FS in charged $B$ decays produce CP asymmetries on the scale of $\sim 5 - 10$  \% for averaged ones and $\sim 50$ \% 
for regional ones.

\noindent
(c) Data show sizable impact of FSI and/or ND in two-body ones and large ones for three-body FS. 

\noindent 
(d) It makes sense that broad (scalar) resonances -- like $f_0(500)/\sigma$ \& $K^*(800)/\kappa$ -- to give large impact on CP asymmetries on three-body FS -- and even more 
for four-body FS, when one goes beyond moments. It means to probe semi-regional CP asymmetries as discussed above in 
Sect. \ref{FOURGENERAL} in general. It depends also on "judgment", which tools one can use. 

\noindent 
(e) I see no reason, why transitions for four-body FS can produce only small CP asymmetries. As again, the SM gives large weak phase 
with $V_{ub}$ in general and $V_{td}$ in special situation; on the other hand the landscapes for strong phases is very complex due to 
FSI as described in Sect. \ref{FOURGENERAL} with one-dimensional observables. Since it depend crucially on strong forces, we cannot 
produce predictions. My point is: I "paint' the situation and focus on correlation with different two- \& four-body FS of $B^0$, $B^+$ and $B_s^0$.

\subsubsection{Comparing with the literature about four-body FS}

CPV in $\Delta B \neq 0$ has been established basically in two-body FS in 2001 - 2007. 
Some theoretical papers about four-body FS have been produced before 2001, but many followed 
the road from Ref.\cite{GERMAN}, namely with special situations $B \to V_aV_b \to h_1h_2h_3h_4$ \cite{ADDL,MLJR}. However, our goals have changed after 2001: we have learnt that the SM gives at least the 
leading source for CPV in $\Delta B \neq 0$. Obviously we have learnt new lessons about the impact 
of QCD forces for exclusive decays. The goal for the future is to find signs of ND as 
a non-leading source for CPV and maybe even its features. It means we cannot focus only on special 
situations as said above; we have to probe four-body FS in general, or at least include broad resonances. 
We cannot quantitatively predict CP asymmetries 
in two- \& four-body FS  ways, but we cannot ignore four-body FS. Again, we cannot trust the lessons we got from diagrams. In particular re-scattering connects U- \& V-spin symmetries. The measured averaged 
triple-product asymmetry in $B^0 \to \phi K^*(892)$ with $\Delta _{CP} = + 0.015 \pm 0.032 \pm 0.005$ 
is consistent with zero CP asymmetry or also non-zero value for direct CPV; furthermore with more data 
we have to probe semi-regional CP asymmetries there and beyond, namely about 
$B^0 \to K^+K^- K^+\pi^-$. 
In general I want to emphasize to probe semi-regional asymmetries in other four-body FS with 
$B_{u,d,s}$. 

Finally, I want to emphasize that the landscapes 
of $\Delta S \neq 0$ and $\Delta B\neq 0$ are quite different. Kaon amplitudes can be described with diagrams based on local operators, where chiral symmetry is a strong tools.  
However, the situations are quite different for $B$ transitions for several reasons. The lessons we have learn from weak kaon decays cannot just transferred.

\subsection{Short resume: Surprises \& maybe some puzzles}

These measured regional asymmetries are much larger than averaged ones in only 
charged three-body FS.  Those show the impact of due to FSI and also much larger than two-body CPV. 
There are informations in several levels; some are expected, while others are more subtle or challenge our understanding of dynamics. 
 
It is very interesting -- but not surprising due to CPT invariance -- that one finds CP asymmetries with opposite signs in 
FS with a pair of $\pi^+\pi^-$ vs. $K^+K^-$ in these three-body FS of $B^{\pm}$ decays. Therefore one 
predicts averaged CP asymmetries with opposite signs $K^0\bar K^0$ vs. 
$\pi^0\pi^0$ (ignoring FS with $\eta$ \& $\eta^{\prime}$ to make it short). 
FSI connect U- and V-spin symmetries in weak decays. 

Well known penguin diagrams $b \Longrightarrow s$ compete well 
(or more) with tree diagram $b \to u \bar u s$ as shown about the rates $B^0 \to K^+\pi^-$ 
vs. $B^0 \to \pi^+\pi^-$ and CP asymmetries. There are several main points. 
\begin{itemize}
\item
The SM gives at least the leading source of measured CPV. Therefore we have to probe CPV for 
signs of ND as non-leading one. 
\item 
The impact of SM penguin diagrams $b \Longrightarrow d$ should be more suppressed than 
$b \Longrightarrow s$; the data about three-body FS 
semi-quantitatively show these rates. 
However averaged CPV already seem to be sizably larger as you can see in 
Eqs.(\ref{SUPP2}, \ref{SUPP6}). 
It describes a quite different situation in three-body FS than in two-body FS.

\item 
Can the SM produce these very large regional ones or not?  
Do they suggest that our control over non-perturbative QCD is  much less than we thought in 
many-body FS? Are there novel lessons about strong forces? 

\item
The best fitted analyses often do not give the best information about underlying dynamics. 
The statements about Eqs.(\ref{SUPP4}, \ref{SUPP8}) depend on the definition about "regional" asymmetries. 
Furthermore we have to include the impact of well-known resonances 
like $f_0(500)/\sigma$ \& $K^*(800)/\kappa$ as long they produce analyses in accepted ways. It is subtle, 
and we have to think about correlations with other transitions. 

\item
Comparing three- \& four-body FS in the weak decays in beauty hadrons at least reveals novel lessons about non-perturbative QCD. 


\item 
Very short comments: 
measuring CP asymmetries in $\Lambda_b^0$ decays gives true challenges for the LHCb collab., and even more for $\Xi_b$ ones. 
Belle II unlikely enter this competition. 

\item
Evidence for CPV in $B^+ \to \bar p p K^+$ was found \cite{LHCbPRL113}. If it stands, it would be 
more than interesting or unusual: large regional CP asymmetry was found like also 
antibaryon-baryon FS. What about $B^+ \to p p \pi^+$ and $B^0 \to \bar p p K^+\pi^-$? Is it a 
novel lesson about strong dynamics about FS or a sign of ND as mentioned before \cite{P209/210}.

\end{itemize}
Furthermore it is crucial to 
include {\em neutral} hadrons in the FS to understand the informations that data give us. 
I have pointed out before like at a recent paper \cite{BUZIOZ}. It was discussed in details in 
Ref.\cite{CHENGHY} based on factorization model. I do not trust those models semi-quantitatively; 
here we pointed out the important impact of re-scattering. 
We know that re-scattering happens all the times and affect sizably asymmetries. 

 It seems that the duality between quarks and hadrons worlds are not close to local 
connections. Other symmetries (and their violations/limits) can limit the classes of hadrons 
involved. It might tell us that we are missing important information about underlying dynamics and have to think more. 
There are good chances that these `prices' will be awarded  with real `prizes'.

\section{$\Delta D \neq 0$ dynamics}
\label{CHARM}

Present data show us that neither indirect nor direct CPV have been found yet in charm hadrons. Finding CP 
asymmetries in charm hadrons -- including charm {\em baryons} -- in the future would be a pioneering achievement. The landscape of data is slim in particular about DCS ones, 
when one talks about CP asymmetries. It will change later from Belle II about $\Delta C \neq 0$ with $D^0$ \& $D^+_{(s)}$ 
and $\Lambda_c^+$ (\& $\Xi_c^{0,+}$) decays.

Non-leptonic FS are given mostly by two-, three- \& four-body FS. 
The impact of CPT invariance is more obvious, but  not trivial. 
The SM produces quite small CPV in SCS transitions  
and basically zero on DCS ones, see Sect.\ref{CKM}. 
We expect non-zero CPV in SCS transitions. 
We can calculate $c \Longrightarrow u$ from intermediate $b$ quark base 
on local operators, and they are tiny; yet intermediate $s$ \& $d$ do {\em not} produced from 
short distance dynamics. Still we `expect' they produce non-zero SCS amplitudes. 

FSI affect DCS amplitudes sizably 
while penguins cannot do it. Furthermore the SM gives  basically zero weak phase there. Thus 
DCS amplitudes give us wonderful places to probe the existence of ND and even its features -- 
if we have enough data. 

There is a well-known example about of the impact of re-scattering and its connection with U-spin symmetry. 
The data show the SCS ratio  
$\frac{\Gamma (D^0 \to K^+K^-)}{\Gamma (D^0 \to \pi^+\pi^-)} \sim \; 3$ 
vs. "originally expected" $\sim 1.4$ based on (broken) U-spin symmetry. 
It was suggested that penguin diagrams might solve this puzzle about two-body FS and shows a `road' for large violation 
of U-spin symmetry \cite{SANDAOLD}. On the other hand one can compare the measured rates of two- and four-body FS 
with only charged hadrons: 
${\rm BR}(D^0 \to \pi^+\pi^-) \simeq 1.4 \times 10^{-3}$ vs.  
${\rm BR}(D^0 \to K^+K^-) \simeq 4 \times 10^{-3}$ and 
${\rm BR}(D^0 \to 2\pi^+2\pi^-) \simeq 7.4 \times 10^{-3}$ vs. 
${\rm BR}(D^0 \to K^+K^-\pi^+\pi^-) \simeq 2.4 \times 10^{-3}$.
Obviously the situation has changed very much from two- to four-body hadronic FS:  
the ratio of two-body FS $\sim 1.4/4\sim 0.35$ changes to the ratio of four-body FS 
$\sim 7.4/2.4 \sim 3.1$ -- i.e., by a factor $\sim$ 10. It is important -- and  not surprising -- 
that FSI has changed the landscapes sizably. However their sum shows:
$\frac{{\rm BR}(D^0 \to K^+K^-) + {\rm BR}(D^0 \to K^+K^-\pi^+\pi^-) }
{{\rm BR}(D^0 \to \pi^+\pi^-) +{\rm BR}(D^0 \to \pi^+\pi^-\pi^+\pi^-) } \sim 0.73$; 
i.e., it is getting close to unity as one expects for inclusive ones in the world of (current) quarks due to $m_d$, $m_s$ being very small on the scale of $\sim $ 1 GeV. 
When one looks at diagrams, it shows the impact of FSI; however, getting numbers is another thing. The situation about CP asymmetries is much more complex with 
weak \& strong phases than about rates; the connection of measured (or measurable) CP asymmetries with hadrons with amplitudes in the world of quarks is subtle. 
However, we cannot ignore four-body FS. 

It has been suggested to probe U-spin symmetry by comparing amplitudes of 
$D^0 \to K^-\pi^+, K^+K^-, \pi^+\pi^-, K^+\pi^-$ \cite{GRONAU}. I quite {\em disagree}:  
effective transition amplitude with re-scattering connect not only charged FS mesons, but neutral ones. 
Furthermore many-body FS have large impact, 
see Sect.\ref{EFFECT}; i.e., the differences between U- \& V-spin symmetries in the world of hadrons are fuzzy, since they connect due to strong forces. 
If the U-spin violations are so small and therefore of the expansion 
of U-spin violations makes sense, I would see that is `luck' so far in a special case -- 
or we miss some important features of non-perturbative QCD. 

Golden \& Grinstein \cite{GOLDEN} were the first to focus on three-body FS in $D$ decays using non-trivial theoretical tools. 
Now we have more refined theoretical tools like dispersion relations. 
Three-body FS with hadrons happen everywhere in the Dalitz plots; interferences appear in different locations. 

Hadronic uncertainties in $c \Longrightarrow u$ decays are discussed in Ref.\cite{BEVMEAD}, in particular about 
$D^0 \to \pi^+\pi^-, \rho^+\rho^-, \rho \pi$. I am somewhat disagree with some of their statements or 
at least their choice of words, namely `tree' ($T$), `$W$-exchanges' ($E$) and 
three `penguin' amplitudes ($P_{d,s,b}$). 
There are subtle, but important differences between diagrams, 
local \& non-local operators; there are non-trivial challenges we have to face as I had said before. Again the left sides of Eqs. (\ref{CPTAMP1},\ref{CPTAMP2}) 
describe amplitudes of hadrons;  the right sides deal with bound states of $\bar q_i q_j$.

First I give a comment or two about CP asymmetries in two-body FS that the 
LHCb collab. presented \cite{NAIK}: 
\bea
A_{CP}(D^+ \to K_SK^+) &=& (0.03 \pm 0.17_{\rm stat} \pm 0.14_{\rm syst} ) \% \\
A_{CP}(D^+_s \to K_S\pi^+) &=& (0.38 \pm 0.46_{\rm stat} \pm 0.17_{\rm syst} ) \% 
\label{LHCbD+Ds+}
\eea
The data are consistent with zero values, but also ${\cal O}(10^{-3})$. One can combine also
\beq
A_{CP}(D^+ \to K_SK^+) + A_{CP}(D^+_s \to K_S\pi^+) = 
(0.41 \pm 0.49_{\rm stat} \pm 0.26_{\rm syst} ) \%
\label{USPIN}
\eeq
Does it show the impact of U-spin symmetry in weak decays? Or does it show the scale of SCS decays at 
${\cal O}(10^{-3})$? What about a combination of both? The LHCb collab. has probed 
$D^0 \to K^+K^-$/$\pi^+\pi^-$ with time-integrated CPV \cite{NAIK}:  
\bea
\nonumber
A_{CP} (D^0 \to K^+K^-) &=& (-0.06 \pm 0.15_{\rm stat} \pm 0.10_{\rm syst}) \% \\
\nonumber
A_{CP} (D^0\to \pi^+\pi^-) &=& (-0.20 \pm 0.19_{\rm stat} \pm 0.10_{\rm syst}) \% \\
\Delta A_{CP} &=& (+ 0.14 \pm 0.16_{\rm stat}  \pm 0.08_{\rm syst}) \%  
\eea
What about the same basic questions as said above: we have to deal with complex situations. 
I emphasize that we cannot focus only on two-body FS. 
In my view the main point is to truly probe CP asymmetries in 
three- \& four-body FS, and we cannot ignore FSI between U- \& V-spin symmetries; i.e., we cannot simply applying U-spin symmetry.  

Actually we know that three- \& four-body FS describe larger parts of the weak decays than two-body ones. Those show a more 
`complex' landscape to combine weak \& strong phases that produce CP asymmetries. Furthermore those data of the rates 
do {\em not} follow the same pattern as we can see from PDG2015: $D^0$ vs. $D^+$ vs. $D_s^+$ for 
SCS \& DCS and 
three- vs. four-body FS. It shows the sizable impact of FSI. We should not be surprised, but we cannot predict theirs even in semi-quantitative ways. 
Finally CPT invariance is `usable' in $D^0$ vs. $\bar D^0$, $D^+$ vs. $D^-$ and $D^+_s$ vs. $D^-_s$  transitions. As said above, it is surprising that it also `usable' for 
$B^+$ vs. $B^-$ decays. Of course, CPT invariance is realized both with charged and neutral pions \& kaons (plus $\eta^{(\prime )})$. 

Again, I use the word of `painting' the FS as the main point: the landscapes for SCS \& DCS decays with 
three- \& four-body FS are complex and different patterns about $D^0$, $D^+$, $D^+_s$ \& $\Lambda^+_c$. It shows the impact of FSI; therefore I  do not discuss the details of the rates.

\subsection{SCS decays of $D^0$, $D^+$ \& $D^+_{s}$ with three- \& four-body FS}
\label{SCS}

PDG 2015 gives rates for three-body FS that can be measured by LHCb in the future, namely mostly with charged hadrons 
in the FS: 
\bea
\nonumber 
{\rm BR} (D^0 \to K^+K_S\pi^-) \sim 2.2 \cdot 10^{-3} &,&
{\rm BR} (D^0 \to K^-K_S\pi^+) \sim 3.6 \cdot 10^{-3} \\
{\rm BR} (D^0 \to K^+K^-\pi^0) \sim 3.4  \cdot 10^{-3} &,&
{\rm BR} (D^0 \to \pi^+\pi^-\pi^0) \sim 14.7  \cdot 10^{-3} \\
\label{D0SCS3}
{\rm BR} (D^+ \to \pi^+\pi^+\pi^-) \sim 3.3 \cdot 10^{-3} &,&
{\rm BR} (D^+ \to \pi^+K^+ K^-) \sim 10 \cdot 10^{-3} \\
\label{D+SCS3}
{\rm BR} (D_s^+ \to K^+\pi^+\pi^-) \sim 6.6 \cdot 10^{-3} &,&
{\rm BR} (D_s^+ \to K^+K^+ K^-) \sim 0.22 \cdot 10^{-3}
\label{Ds+SCS3}
\eea
I have no predictions for these rates or quantitative one for CP asymmetries beyond saying that averaged 
CP asymmetry are of the level of ${\cal O}(0.1)$\% from the SM.  In the future we have to probe Dalitz plots 
with the impact of FSI on regional CP asymmetries and their correlations due to CPT invariance 
\footnote{Of course these correlations can be satisfied in other FS with neutral mesons.}. 
It was discussed in 
Ref.\cite{REIS} with simulations of $D^{\pm} \to \pi^{\pm} \pi^+\pi^-$ and 
$D^{\pm} \to \pi^{\pm} K^+K^-$ with small weak phases and sizable resonances phases in the world of 
hadrons. 

Here I want to emphasize that we see quite different landscapes for charm mesons 
mostly with charged FS. For $D^0$ decays one sees that the combination 
of a pair of $\bar KK$ vs. without them are somewhat similar, while these pair larger than without them for $D^+$; 
on the other hand $D^+_s$ decays will us much smaller rates with a pair of $\bar KK$ than without. 
Those show the impact of FSI in different ways. There are 
good reasons why to compare binned "fractional asymmetries'" vs. "significance" \cite{REIS} vs. "un-binned" ones \cite{WILL}. 
A short comment: it might show the different impact of WA on rates and effective phases for $D^+$ vs $D^+_s$.

Again, it should not be the final step. Subtle, but powerful tools have not been applied yet like 
dispersion relations \cite{DR,KUBIS,KUBIS2}; those are based on low energy collisions between two hadrons 
in the future. We have the tools to do that -- but it takes time. It is always wonderful to get more ideas, 
but there is no reason to wait for novel ideas. 

I add a comment that connect with FSI: LHCb has measured SCS $D^0 \to \pi^+\pi^- \pi^0$ \cite{3PIONS}.   
Their paper starts with "The decay $D^0 \to \pi^- \pi^+\pi^0$ ... proceeds via a singly Cabibbo suppressed 
$c\to d u \bar d$ transition with a possible admixture from a penguin amplitude." 
It seems to ignore FSI from $c\to s u \bar s$ diagram; I see no good reason for that. Above in Sect.\ref{EFFECT} 
I have discussed the impact of re-scattering in general; here we cannot ignore refined local operators $c\to s u \bar s$. 
Of course, one 
measure the rates \& asymmetries of $D^0 \to 3 \pi$ \& $D^0 \to \pi \bar K K$ in the world of hadrons. My point 
is to connect amplitudes in the world of quarks. Duality is not trivial at all. 

Data above show that $\Gamma (D^0 \to \pi^+\pi^-\pi^0)$ is larger than single $\Gamma (D^0 \to \pi \bar K K)$; however we get 
$\sum \Gamma (D^0 \to \pi \bar K K) \sim \Gamma (D^0 \to 3 \pi)$.  

Focusing on four-body FS with at most one neutral mesons I list FS that LHCb collaboration can measure. 
\bea
{\rm BR} (D^0 \to 2\pi^+ 2\pi^-) \sim 7.5 \cdot 10^{-3}    &,&
{\rm BR} (D^0 \to K^+ K^- \pi^+\pi^-) \sim 2.4 \cdot 10^{-3}  \\
\nonumber
{\rm BR} (D^+ \to K^+K_S \pi^+\pi^-) \sim 1.7 \cdot 10^{-3}  &,& 
{\rm BR} (D^+ \to K^- K_S \pi^+\pi^+) \sim 2.3 \cdot 10^{-3} \\
{\rm BR} (D^+ \to \pi^+ \pi^- \pi^+\pi^0) &\sim & 11.7 \cdot 10^{-3}  \\
{\rm BR} (D^+_s \to K_S \pi^+ \pi^- \pi^+) &=& (3.0 \pm 1.1)\cdot 10^{-3}   
\eea
The pattern is quite different from the three-body FS: four pions produce the leading source so far. 

T-odd momenta has been measured in $D^0 \to K^+K^-\pi^+\pi^-$ \cite{DTOKKPIPI}: 
\bea
\nonumber 
A_T = (-7.18 \pm 0.41 \pm 0.13 ) \cdot 10^{-3}  \; &, &\; 
\bar A_T = (-7.55 \pm 0.41 \pm 0.12 ) \cdot 10^{-3} \\
a_{CPV}^{\rm T-odd} &=& (0.18 \pm 0.29 \pm 0.04) \%
\eea
It shows sizable impact of FSI: $A_T \neq 0 \neq \bar A_T$. On the other hand, 
no CP violation has been found yet. From these data I learn, namely that is just the beginning: 
\begin{itemize}
\item
One has to probe semi-regional CPV in $D^0 \to K^+K^-\pi^+\pi^-$ with different tools as discussed above in Sect. \ref{FOURGENERAL} in general. 

\item
One can follow the somewhat roads for $D^0 \to 2 \pi^+ 2\pi^-$. There is a complication, namely how 
to differentiate between 2 $\pi^{\pm}$. 

\item 
With more data one can think about comparing the informations from $D^0 \to K^+K^-\pi^+\pi^-$ vs. 
$D^0 \to 2 \pi^+ 2\pi^-$ based on CPT invariance. 

\item
We have to follow the same `roads' in smart ways in $D^+ \to K_S\pi^+K^{\pm}\pi^{\mp}$ and 
$D_s^+ \to K^+\pi^+\pi^-\pi^0$/$K^+K^+K^-\pi^0$/$K_S\pi^+\pi^-\pi^+$/
$K_SK^+K^-\pi^+$. 

\end{itemize}
Again, I want to emphasize that drawing penguin diagrams is one thing, but trust the numbers they give is quite different.

\subsection{DCS decays of $D^0$, $D^+$ \& $D^+_s$ with three- \& four-body FS}
\label{DCS}

For theorists the landscapes are much simpler for the impact of ND, since the SM background is close 
to zero with $c \to u d \bar s$ with $\Delta C =1$ \& $\Delta S = - 1$. Furthermore there are only 
two refined tree operators, and penguin diagrams cannot contribute; however 
re-scattering does contribute sizably.  Of course we need huge numbers of charm hadrons.

Present data from PDG15 are even slimmer: 
\beq
{\rm BR}(D^0 \to K^+\pi^-) \sim 1.5  \cdot 10^{-4}
\eeq
Two-body FS sets the scale. 
\bea 
{\rm BR} (D^0 \to K^+ \pi^- \pi^0) & \sim & 3.1 \cdot 10^{-4} \\
{\rm BR} (D^+ \to K^+\pi^+\pi^-) \sim 5.5 \cdot 10^{-4} &,&
{\rm BR} (D^+ \to 2 K^+ K^-) \sim 0.90  \cdot 10^{-4} \\ 
{\rm BR}(D^+_s \to 2K^+\pi^-)&\sim & 1.3 \cdot 10^{-4}
\eea

So far there are no real surprise in the landscape; of course there are sizable experimental uncertainties. 
Impact of ND can hide in $D^0 \to K^0 \pi \pi$ due to Cabibbo favored $D^0 \to \bar K^0\pi \pi$ transitions. 

On the other hand CP asymmetries can be probed in these $D^+\to K^+\pi^+\pi^-/K^+K^+K^-$ and  
$D_s^+ \to K^+K^+\pi^-$, where the SM gives hardly any background; of course, we need more data. 
In the future we can measure "fractional asymmetry", "significance" or "unbinned" Dalitz plot \cite{MIR1,REIS,WILL}.

So far we have virginal landscape for four-body FS, except:
\beq
{\rm BR}(D^0 \to K^+\pi^+2\pi^-) \sim 2.6 \cdot 10^{-4} 
\eeq
LHCb should be able to measure averaged and practical definitions of regional CP asymmetries 
as discussed in Sect.\ref{FOURGENERAL} in $D^+ \to K^+\pi^-\pi^+\pi^0, K^+K^-K^+\pi^0$, $K_S\pi^+\pi^+\pi^-$ and 
$D^+_s \to K^+K^+\pi^-\pi^0$, $K^+K_S \pi^+\pi^-$, $K^+K^+\pi^-\eta$  for second steps.

\subsection{General comments about CP asymmetries in charm baryons}
\label{CHBAR}

Charm baryons might  turn out to be the `Poor Princesses' \cite{TIM}
for establishing CPV in baryons' decays and showing the impact of ND there. It is 
very important to probe correlations between $\Lambda_c^+$, $\Xi_c^+$ 
\& $\Xi_c^0$ decays in different levels 
\footnote{When one understands the political landscape in Bavaria and the rest of Germany, one 
might see the analogy of $\Xi_c^+ = [csu]$ vs. $\Lambda_c^+=[cdu]$.}. 
For DCS one gets WA diagram $[cs] \Rightarrow du$, 
while $[cd] \Rightarrow du$ and $[cs] \Rightarrow  s u$ for SCS. Their impacts depend on 
their wave functions.

It is not surprising that the Cabibbo favored  $\Lambda_c^+ \to p K^-\pi^+$ decay produces sizable branching ratio: 
\beq
\nonumber 
{\rm BR} (\Lambda_c^+ \to p K^-\pi^+) = (6.84^{+0.32}_{-0.40}) \cdot 10^{-2} 
\eeq
LHCb will be able to measure this with accuracy soon; it can be used to calibrate branching ratios. Finding CP 
asymmetry there, it would be a miracle. To be realistic: one probes production asymmetry in pp collisions  
by measuring the ratio of $\bar \Lambda_c ^- \to \bar p K^+\pi^-$ vs. $\Lambda_c^+ \to p K^-\pi^+$
\footnote{LHCb might be able also measure $\Lambda_c^+ \to p \bar K^0 \Rightarrow p K_S $ with accuracy. We know that indirect CP violation in $K^0 - \bar K^0$ 
oscillations has been measured by Re $\epsilon_K$; it checks experimental (un)certainty.}. Thus one can calibrate the branching ratios of suppressed decays, when one probes CP asymmetries as 
discussed below. 

The landscape is slim, when one wants to describe it in a `positive' way: 
\bea
{\rm BR}(\Lambda_c^+ \to p \pi^+\pi^-) &=& (4.7 \pm 2.5) \cdot 10^{-3}\\
{\rm BR}(\Lambda_c^+ \to p K^+ K^-) &=& (1.1 \pm 0.4) \cdot 10^{-3}
\eea
It is not clear what we can learn from these numbers about underlying dynamics. 
Small CP asymmetries are most likely to appear {\em both} in 
$\Lambda_c^+ \to p \pi^+\pi^-$ with $p\rho^0$, $p \sigma$ etc. and 
$\Lambda_c^+ \to p K^+K^-$ with $p\phi$, $\Lambda^*K$ etc. 
As before, regional CP asymmetries are usually much larger than averaged asymmetries in 
three-body final states. 

As the second step we have to probe Dalitz plots in `model insensitive' ways, but again not as a final 
step. One has to be realistic: to extract the dynamics information underlying the finite data one has 
to apply quantitative theoretical analyses based on "judgment".

The present limits for these DCS decays are in the region one expects: 
\beq
{\rm BR} (\Lambda_c^+ \to p K^+\pi^-) < 3.1 \cdot 10^{-4} \; .
\eeq

One has to remember that the landscapes of $\Lambda_c^+ \to p K^-\pi^+$ 
vs. $\Lambda_c^+ \to p K^+\pi^-$ are quite different due to contributions from both narrow \& broad resonances.  
It is less complex than for SCS ones, since they are described by two refined local operators without penguin diagrams. 
Beyond the need for large numbers of $\Lambda_c^+$ the main challenges is to deal with production asymmetries in 
$pp$ collisions. 

As mentioned above, SM produces only one quark amplitude for DCS transitions; therefore SM cannot produce CP asymmetry. 
Furthermore the size of SM amplitudes are very much suppressed; 
thus it gives more sensitivity to the impact of ND.  

In particular, one can analyze 
$\Lambda_c^+ \to p K^+\pi^-$ vs. $\bar \Lambda_c^- \to \bar p K^-\pi^+$  for CP violation and compare with CF 
$\Lambda_c^+ \to p K^-\pi^+$ \& $\bar \Lambda_c^- \to \bar p K^+\pi^-$ to learn about the impact of FSI. Of course, one has to differentiate $K^+\pi^-$ from 
$K^-\pi^+$ in the $\Lambda^+_c$ decays despite the huge difference in their branching ratios. 
These DCS decays have not been found yet. On the other hand one can hope for significant CP violation in the DCS transitions.  

The $\Lambda_c$ final states include $pK^*$, $p\kappa$, $N^*K$, $\Delta ^{(*)}K$, 
$\Lambda^*\pi$ etc.  -- i.e., numerous states that give us lessons about the existence of ND 
and its features due to several reasons.  
These are qualitative and at most a semi-quantitative comments. Quantitative theoretical works 
will happen based on dispersion relations, but they will take more efforts and time 
(in particular for $pK$ \& $p\pi$ states). 

Of course crucial jobs will be done by Belle II, and more refined theoretical technologies have to be applied to understand the data about the underlying dynamics.

\section{Summary}
\label{SUM}

I talk mostly about strategies in heavy flavor dynamics and discuss the landscapes.  
\begin{itemize}
\item 
We know that the SM gives us at least the leading source of CPV in the transitions of beauty mesons 
(and $K_L \to \pi \pi$). 
Therefore we have to probe those transitions about non-leading sources; at least we learn about 
non-perturbative QCD. 

\item
It is very important to establish CPV in charm hadrons. Furthermore we expect ${\cal O}(0.001)$ averaged 
CP asymmetries in SCS decays and basically zero in DCS ones from the SM. It is due to new parametrization  
of the CKM matrix \cite{AHN}, which is less obvious, but more consistent. DCS amplitudes might be an excellent hunting for ND -- 
if we have enough data and apply refined analyses.

\item 
To find CP asymmetries in beauty \& charm {\em baryons} is an important achievement. At least it gives novel lessons about strong dynamics. 

\item 
On the theoretical side no large progress has been established in understanding 
fundamental dynamics in many-body FS. On the other hand 
data from the LHCb collaboration give theorists excellent reason to enter the `game'. 
There are two different `cultures', namely Hadrodynamics and HEP. It is not trivial to combine their tools; 
however, it is crucial to apply in new landscapes with accuracy.

\item 
I see {\em no reason} why two-, three- \& four-body FS follow the same pattern in the decays both of 
beauty \& charm hadrons. Actually data show different patterns for rates and CP asymmetries. Based on 
experience we know that FSI/re-scattering has impact, in particular  about connections with U- \& V-spin symmetries.

\item
It is important to measure {\em correlations} between charm and beauty hadrons (\& $\tau$ leptons), 
{\em not} focus on one or two `golden' tests. 

\item 
It is not a good idea to base our conclusions coming from the best fitting of the data 
available; we need deeper thinking 
\footnote{It was suggested by Stephan Ellis from the University of Washington in a 
very different situation, namely to probe internal structures of jets with "pruning" \cite{PRUNING}; 
however the main point is the same, namely "judgment".}.

\item 
We have to probe regional CP asymmetries in many-body FS, namely three- \& four-body FS.  
It is an excellent achievement by the LHCb by measuring  
regional CP asymmetries in $B^- \to K^- \pi^+\pi^-$/$K^-K^+K^-$/$\pi^-\pi^+\pi^-$/
$\pi^-K^+K^-$ on the experimental side. 
We have to use tested tools based on dispersion 
relations etc., which shows collaborations of experimenters \& theorists backward \& forward 
\footnote{One can learn from the history of art: "It is better to imitate ancient than modern work." 
Leonardo da Vinci(1452 - 1519).}.

\item
The impacts of penguin diagrams are complex. In the SM one predicts inclusive {\em beauty} decays. 
For exclusive ones we need help from other tools like chiral symmetry \& dispersion relations.  
There are other problems about penguin diagrams in charm transitions. They give pictorials 
for SCS ones including FSI, but not much more; they cannot produce DCS amplitudes. 

\item 
The measured (or measurable) rates and their CP asymmetries are shown on the left side of 
Eq.(\ref{REGCPV}). On the right 
side it shows the amplitudes that are based on the theory of the SM and/or ND (assuming CPT invariance). 
The connections of the worlds of hadrons vs. quarks (\& gluons) are complex; i.e., the connections of 
the diagrams with underlying dynamics are not straightforward -- namely the impact of 
re-scattering $ a \to b + c$, whether one uses amplitudes of hadrons or quarks. 

\end{itemize}
Here I had focused on three- \& four-body FS with only charged ones and maybe also with one neutral hadron, where 
LHCb has measured and will continue. 
It is very important to probe FS with more neutral ones as Belle II will do it -- but it will not happen very soon.

\vspace{0.5cm}

{\bf Acknowledgments:} This work was supported by the NSF under the grant numbers PHY-1215979 
\& PHY-1520966. 

\vspace{4mm}


\end{document}